\newcommand{\draft}{F}

\if T\draft
\documentclass[11pt,twoside,draft]{IEEEtran}
\usepackage{psfig}
\else
\documentstyle[psfig,epsf,twocolumn,twoside]{IEEEtran} 
\fi

\newcommand{\briefrefs}{T}

\if T\briefrefs

\else

\fi

\newcommand{\beqn}{\begin{equation}}
\newcommand{\eeqn}{\end{equation}}

\newcommand{\be}{\begin{equation}}
\newcommand{\ee}{\end{equation}}
\newcommand{\beqa}{\begin{eqnarray}}
\newcommand{\eeqa}{\end{eqnarray}}
\newenvironment{vectra}[1]%
	{\left[ 		
	\begin{array}{#1}}
	{\end{array} \right]}

\newcommand{\bv}{\begin{vectra}}
\newcommand{\ev}{\end{vectra}}

\newcommand{\ba}{\begin{array}}
\newcommand{\ea}{\end{array}}

\newcommand{\define}{\stackrel{\triangle}{=}}

\newcommand{\Comment}[1]{}

\newcommand{\figureLongCapTwoCol}[3]{
	\if F\draft
	\begin{figure}[htbp]
	\centerline{\psfig{figure=#2,width=3.25in}} 
	\caption{\protect {\small #3}}
	\label{#1} 
	\end{figure}
	\fi
}

\newcommand{\Roc}{F}
\newcommand{\Long}{T}
\newcommand{\Short}{F}

\if T\Roc
\newcommand{\roc}[1]{#1} 
\else
\newcommand{\roc}[1]{} %
\fi
\if T\Long
\newcommand{\lungo}[1]{#1} 
\else
\newcommand{\lungo}[1]{} %
\fi
\if T\Short
\newcommand{\corto}[1]{#1} 
\else
\newcommand{\corto}[1]{} %
\fi

\begin{document}
\bibliographystyle{IEEE}

\title{Fractionally-addressed delay lines}
\author{Davide Rocchesso\thanks{D. Rocchesso is with the Universit\`a di Verona - Dipartimento Scientifico e Tecnologico,  Strada Le Grazie, 15 - 37134 Verona, Italy, E-mail: rocchesso@sci.univr.it. 
Phone: ++39.045.8027979, FAX: ++39.045.8027982. 
~~~{\copyright 2000 IEEE. Personal use of this material is permitted. However, permission to
                          reprint/republish this material for advertising or promotional purposes or for creating new
                          collective works for resale or redistribution to servers or lists, or to reuse any copyrighted
                          component of this work in other works must be obtained from the IEEE.}
 }}
\markboth{IEEE Trans. on Speech and Audio, 2000. Accepted for publication.}{Rocchesso: {Fractionally-addressed delay lines} }

\date{\today}

\maketitle
\begin{abstract}
While traditional implementations of variable-length digital delay lines are based on a circular buffer accessed by two pointers, we propose an implementation where a single fractional pointer is used both for read and write operations. 
\Comment{A novel realization of the digital delay line is
proposed. It uses a memory buffer with a single fractional
pointer rather than a reading and a writing pointer. }
On modern general-purpose architectures, the proposed method is nearly as efficient as the popular
interpolated circular buffer, and it behaves well for delay-length modulations commonly found in digital audio effects. 
The physical interpretation of the new implementation shows that it is suitable for simulating tension or density modulations in  wave-propagating media.
\end{abstract}


\section{Introduction}
The digital delay line is a fundamental component of many signal processing architectures in several application fields. 
In audio signal processing, delay lines are used to implement audio effects, such as reverberation or pitch shifting, or to model wave propagation in physical models of musical instruments. 

Since the early days of computer music, the delay line has been proposed as a building block in software synthesis languages such as Music V~\cite{roads}. In those implementations, a fixed delay of an integer number $D$ of samples was implemented by means of a circular queue having length $D$. At each time sample,  a read and then a write operations were performed at the location pointed by the single circulating pointer\footnote{This is the way the {\tt delay} unit generator is still implemented in the popular sound processing language Csound~\cite{roads}.}. In order to have fractional lengths, linear interpolation between adjacent memory locations right behind the pointer was introduced.

Since dynamic variations of the delay length are required by many important applications, such as pitch shifting, variable delay lines were introduced in several signal processing environments and languages.
The classic implementation of the variable-length digital delay line uses a
circular buffer, which is accessed by a writing pointer followed by a reading
pointer\footnote{This is the way the {\tt vdelay} unit generator is implemented in Csound.}~\cite{Orfanidis}.
When the delay length has to be made variable, the relative distance
between the reading pointer and the writing pointer is varied sample
by sample. In order to allow for fractional lengths and click-free
length modulation, some form of interpolation has to be applied at the
reading point~\cite{Zolzer,LaaksoB,TassartDepalle97}.
The following properties should be ensured by the interpolation
device: \corto{(i) flat frequency response, (ii) linear phase response, (iii) transient-free response to variations of the delay length.}
\lungo{
\begin{itemize}
\item[1.] flat magnitude frequency response
\item[2.] linear phase response
\item[3.] transient-free response to variations of the delay.
\end{itemize}}
\lungo{FIR filters, usually in the form of Lagrange interpolators~\cite{LaaksoB,Zolzer}, are widely
used. Even though they can not satisfy property 1, they are certainly
compliant to property 3 and can also satisfy property 2 to a great
extent on a wide
frequency range~\cite{LaaksoB}.
On the other hand, allpass filters, often designed to have a maximally-flat delay
response at low frequencies, satisfy property 1 exactly but are quite
nonlinear in their phase response~\cite{LaaksoB}. Moreover, a rather complicated
structure has to be devised in order to attain property 3 by means of allpass filters~\cite{VesaT,VesaSP98}.}

All of the prior realizations, as far as a fixed delay length is
considered, are linear and time-invariant systems, thus being
completely described by their frequency response.
Vice versa, we are proposing a realization which is time-varying even
in the case of constant delay. This realization was first proposed and implemented by the author as part of a thesis work~\cite{laurea}. Afterwards, improved implementations and input-output analyses were sparsely presented at some conferences~\cite{roccim98,RocchessoDAFX98}. This paper  systematizes the main ideas and results, and gives some hints to use the technique in musical applications.

In section~\ref{lengthmod} we recall the classic FIR realization of the delay line and show the magnitude and phase responses in the case of Lagrange quadratic interpolation. In particular, we show how length modulation affects the spectrum of an incoming signal.
In section~\ref{fadline} we describe the novel Fractionally-Addressed Delay (FAD) line, and in section~\ref{io} we analyze its input-output behavior in terms of signal-to-error ratio and transient response to delay variations.  
In section~\ref{physical} we interpret the FIR and FAD lines as physical wave-propagating media, and we illustrate the application of the FAD line in physical modeling of resonators.
In section~\ref{performance} the FIR and FAD lines are compared in terms of performance in general-purpose superscalar architectures, and the convenience of using FAD lines in digital audio effects is briefly discussed. 

\section{Length-Modulated Delay Lines}
\label{lengthmod}

Nowadays, a plethora of techniques is available to design FIR filters to be used as interpolators in fractional delay lines~\cite{LaaksoB}. Such interpolators can be designed in order to be optimal in one of several senses. However, most of the optimal FIR interpolators are based on sets of precomputed coefficients or use windowing to compute the coefficients online. On the other hand, the coefficients of Lagrange interpolators are easy to compute and the resulting filter has the desirable property of having maximally-flat magnitude in low frequency. Moreover, the magnitude response is guaranteed to be less than unity at any frequency, and this is a fundamental property if the fractional delays have to be used within feedback structures, such as physical models. For these reasons, Lagrange-interpolated delay lines are very popular in sound synthesis, digital audio effects, and wherever the delay length has to be modulated at runtime. In this section, we consider the Lagrange interpolator as a reference case, focusing on the second order FIR filter used e.g., in~\cite{ValimakiICMC98}. We introduce an approximate analysis of the length-modulated delay that will be useful to understand the dynamic behavior of the new FAD line.

\subsection{Lagrange-interpolated delay lines}
A Lagrange interpolator can be characterized at any fractional delay by its magnitude and phase delay responses~\cite{LaaksoB}. At any frequency $f$ and for a given interval of fractional delays, the interpolator has a magnitude response ranging from $A_{\rm min}$ to $A_{\rm max}$ and a phase delay ranging from  $\tau_{\rm min}$ to $\tau_{\rm max}$. For example, a second-order (quadratic) interpolator can be constructed by using the coefficients~\cite{VesaT}
\beqa
h_0 & = & d (1 + d) /2 \nonumber \\
h_1 & = & (1 + d) (1 - d) \\
h_2 & = & -d (1 - d) /2 \nonumber
\eeqa 
in the FIR transfer function $H(z) = h_0 + h_1 z^{-1} + h_2 z^{-2}$.
If we enforce $d \in \left[-1.0 \dots 1.0 \right]$, the magnitude is constrained to be less than one, and the phase delay deviates roughly sinusoidally around a straight line going from $0$ to $2$ samples. Figures~\ref{quaddel}.a and~\ref{quaddel}b show the magnitude and excess phase-delay responses at $f=F_s/4$ as functions of $D \define 1 - d$. Ideally, one would like the excess phase delay to be zero, in such a way that the ideal phase delay is equal to $D$. 
\if F\draft
\begin{figure}[ht]
\psfull
\psfig{file=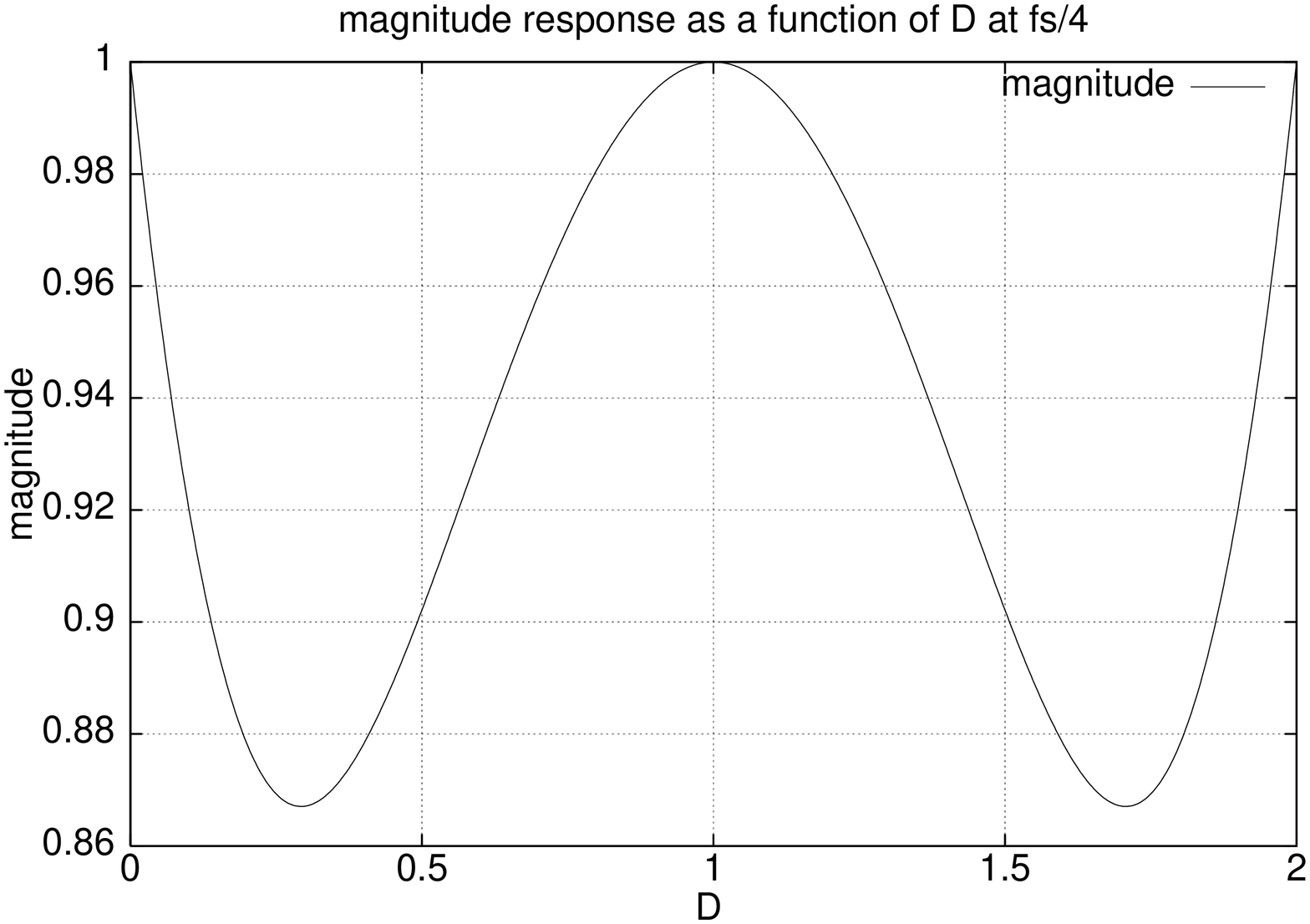,width=8.0cm,angle=-90}
\centerline{(a)}
\vspace{0.1cm}
\psfig{file=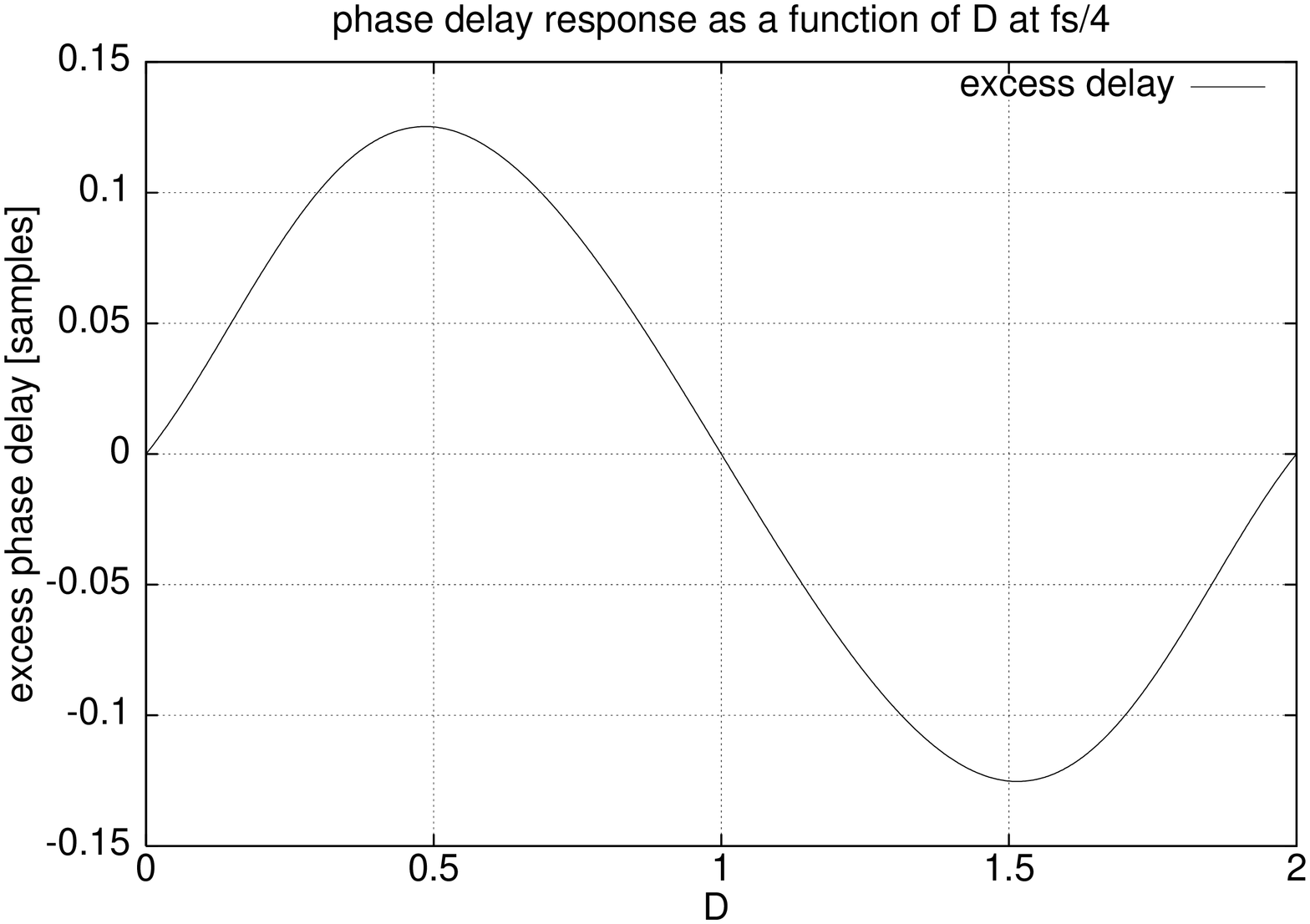,width=8.0cm,angle=-90}
\centerline{(b)}
\caption{Magnitude (a) and excess phase-delay (b) responses for a quadratic Lagrange interpolator as functions of the parameter $D = 1 - d$ at frequency $f=F_s/4$}
\label{quaddel}
\end{figure}
\fi

\subsection{Approximate analysis of length-modulated delay lines}
If the parameter $d$ is continuously varied around $0$, it is clear that frequency-dependent amplitude and phase modulations are both applied to the signal. 
In order to understand how these modulations affect the spectrum, we assume that the magnitude and phase delay vary sinusoidally\Comment{ with the parameter $d$}\footnote{This is quite a strong assumption, but the goal of this analysis is just to outline a qualitative behavior of the spectrum. The actual behavior will be slightly different, especially because of deviations from zero-mean and sinusoidal variation of $d$.\Comment{due to deviations from sinusoidal modulation and from zero-mean values of $d$.}}, and that the frequency of the magnitude sine is twice the frequency $\omega_M$ of the phase-delay sine. This latter assumption is justified by the aspect of figure~\ref{quaddel}, where there are two minima in figure~\ref{quaddel}.a and one minimum in figure~\ref{quaddel}.b.
If the input is  a cosine wave  at frequency $\omega_0$, the output signal takes form
\be
v_d = A_m\left(1 + m \cos{2 \omega_M t} \right) \cos{\left(\omega_0 t + \tau_{\rm max}\omega_0 \sin{\omega_M t}\right)} \,,
\ee
where
\be
m = \frac{1 - A_{\rm min}}{1 + A_{\rm min}} \,,
\ee
\be
A_m = \frac{1 + A_{\rm min}}{2} \,,
\ee
and $A_{\rm min}$ and $\tau_{\rm max}$ are the minimum magnitude and maximum delay for the given interval of variability of $d$ and frequency $\omega_0$.
If we enforce $d \in \left[-0.5 \dots 0.5 \right]$, the extremal magnitude and phase delay responses as functions of frequency are plotted in figures~\ref{quadmagdel}.a and~\ref{quadmagdel}.b, respectively.
\if F\draft
\begin{figure}[ht]
\psfig{file=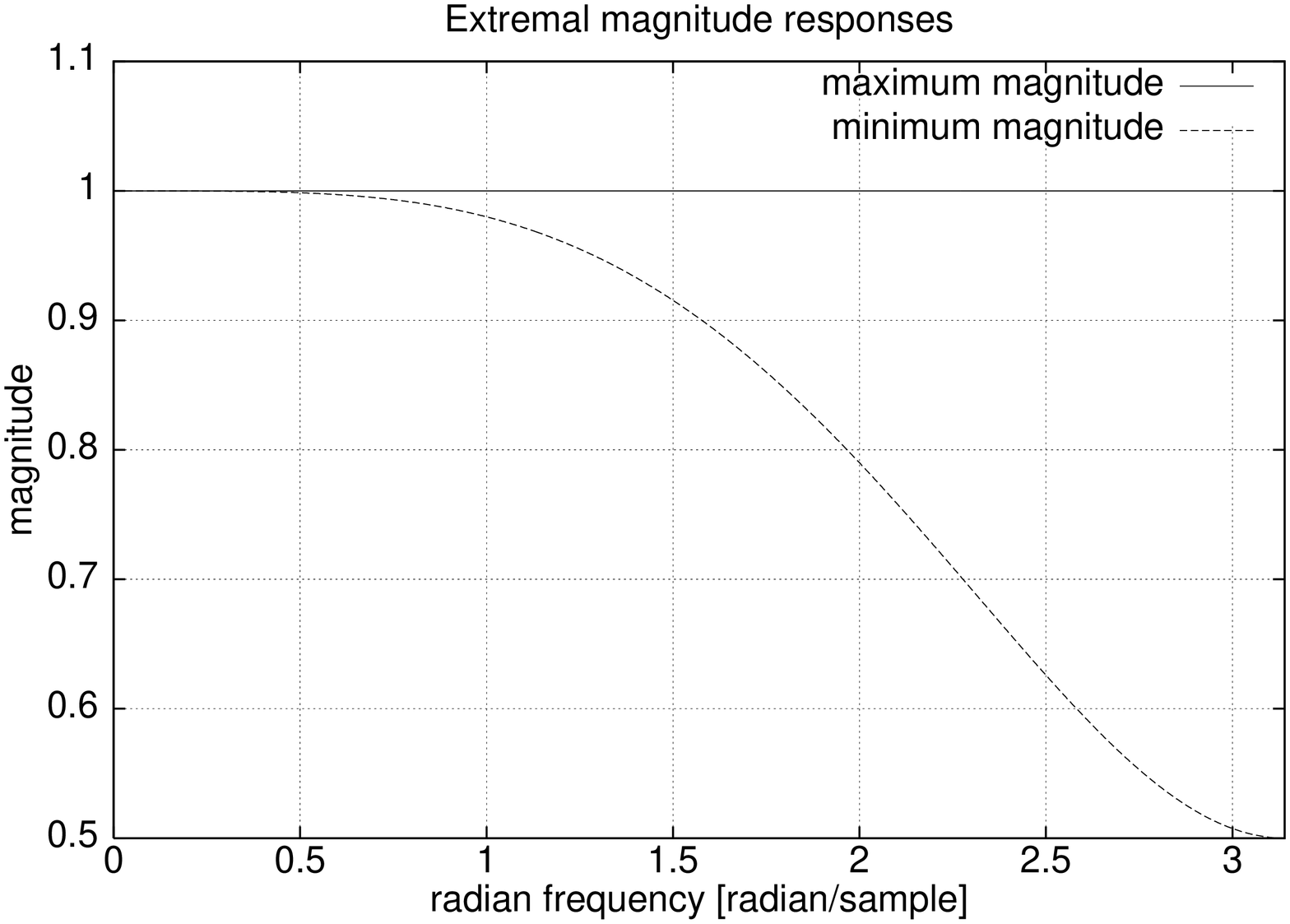,width=8.0cm,angle=-90}
\centerline{(a)}
\vspace{0.1cm}
\psfig{file=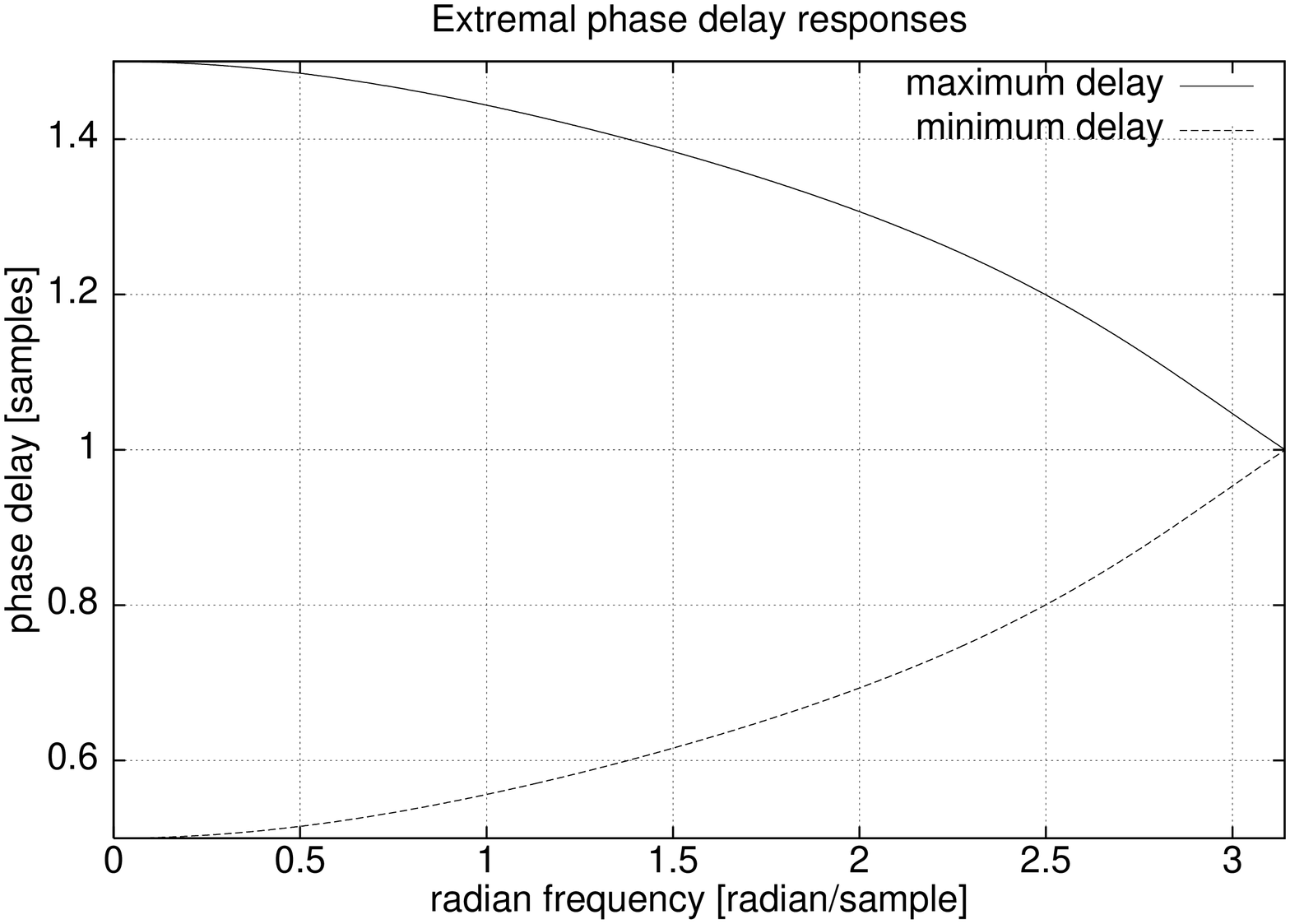,width=8.0cm,angle=-90}
\centerline{(b)}
\caption{Extremal magnitude (a) and phase delay (b) responses for a quadratic Lagrange interpolator}
\label{quadmagdel}
\end{figure}
\fi

The amplitude modulation gives rise to a carrier and two side bands:
\beqa
v_d & = & A_m \cos{\left(\omega_0 t + \tau_{\rm max}\omega_0 \sin{\omega_M t}\right)} \\
   &  & +\frac{m A_m}{2}\cos{\left((\omega_0 - 2\omega_M)t + \tau_{\rm max}\omega_0 \sin\omega_M t  \right)} \nonumber\\
   &  & + \frac{m A_m}{2}\cos{\left((\omega_0 + 2\omega_M)t + \tau_{\rm max}\omega_0 \sin\omega_M t  \right)} \; .\nonumber 
\eeqa
The phase modulation generates infinitely many sidebands of the three components of amplitude modulation. However, if we assume that the sidebands of order higher than two are negligible~\footnote{This assumption is justified by experimental observations.}, the resulting signal can be expressed as
\beqa
v_d & = & A_0 \cos{\omega_0 t} \\
& & - A_1 \cos{(\omega_0 - \omega_M) t} + A_1 \cos{(\omega_0 + \omega_M) t} \nonumber \\
& & + A_2 \cos{(\omega_0 - 2 \omega_M) t} + A_2 \cos{(\omega_0 + 2 \omega_M) t} \nonumber \\
& & + \dots  \; , \nonumber
\eeqa
where the coefficients $A_i$ are given by table~\ref{modprod}
and $J_i$ is the $i$-th order Bessel function of the first kind evaluated in $ \tau_{\rm max}\omega_0$.
\if F\draft
\begin{table}[h]
\begin{tabular}{|c|c|c|} \hline
$A_0$ & $A_1$ & $A_2$ \\ \hline\hline 
& & \\
$A_m J_0 + {m A_m} J_2 $ & $ A_m J_1 - \displaystyle{\frac{m A_m}{2} J_1} $ & $ \displaystyle{A_m J_2 + \frac{m A_m}{2} J_0} $ \\ 
& & \\ \hline
\end{tabular}
\caption{Amplitude of the sidebands of an amplitude- and phase-modulated signal}
\label{modprod}
\end{table}
\fi

Figure~\ref{components} shows the magnitude of the carrier and two side modulation products as a function of frequency, where $D=1$ ($d=0$) and the sinusoidal length modulator has amplitude $0.5$ (i.e. it modulates the central part of figure~\ref{quaddel}).
\if F\draft
\begin{figure}[hbt]
\psfig{file=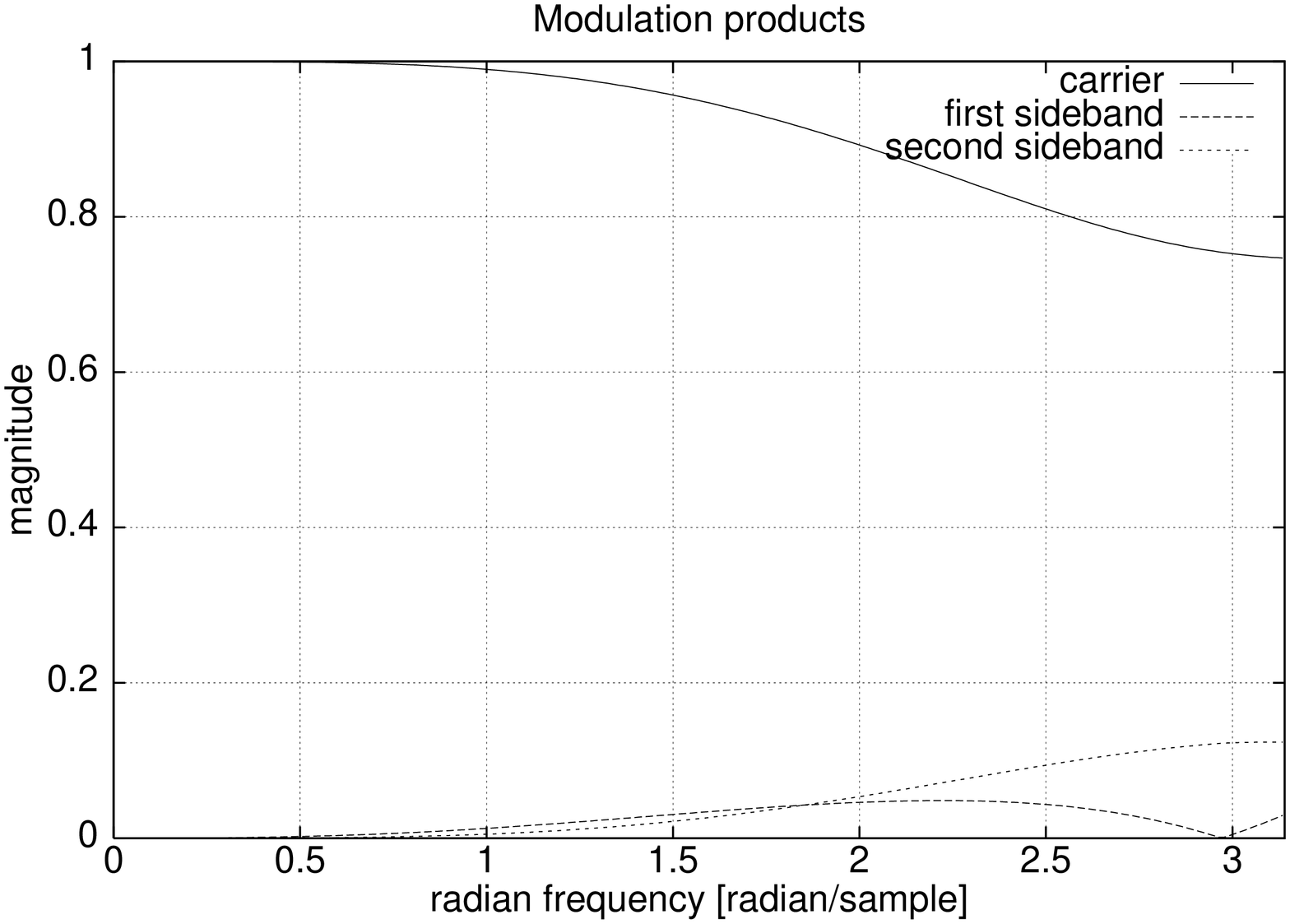,width=8.0cm,angle=-90}
\caption{Carrier and side modulation products as a function of carrier frequency }
\label{components}
\end{figure}
\fi
\subsection{Delay-length dithering}
The general trend expressed by figure~\ref{components} is that modulation acts as a lowpass filter on the carrier and as a kind of highpass filter for the side components. 
An interesting point is made by moving the center of modulation slightly away from $0$ and observing that
we get output signal components that are very similar to those of figure~\ref{components}. On the other hand, a fixed-length interpolated delay would exhibit different magnitude responses at different values of $d$, in a range bounded by the curves of figure~\ref{quadmagdel}.a. As a consequence, we can use delay-length modulation in order to have a more uniform frequency response for different values of $d$, as it is often required in applications such as physical modeling (see section~\ref{physical}). In other words, small-range modulation can be used as a sort of dithering in the delay length, and possibly put on top of long-range modulations. 
As an example, figure~\ref{noisyramp} shows the sonogram of the response of a non-modulated and modulated FIR line to a pulse train. In both cases, the delay length is slowly linearly increased in order to explore different fractional values of $d$. In figure~\ref{noisyramp}.a the valleys corresponding to the minimum-magnitude curve of figure~\ref{quadmagdel}.a are visible as darker areas. In figure~\ref{quadmagdel}.b the darker areas are more uniformly spread along the horizontal axis, thus allowing magnitude compensation by fixed high-pass filtering.
\if F\draft
\begin{figure}[hbt]
\centerline{\psfig{file=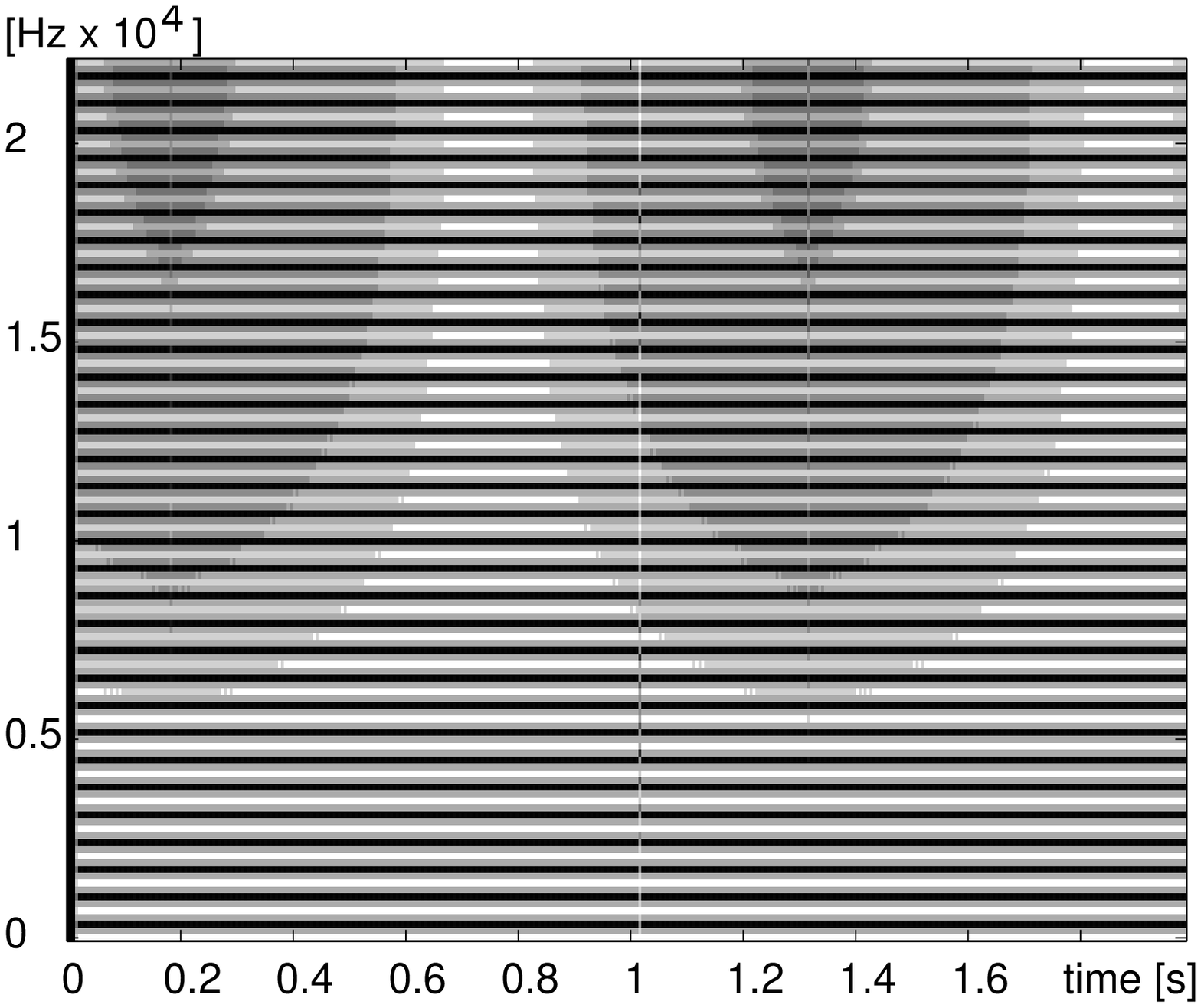,width=8.0cm}}
\centerline{(a)}
\vspace{0.1cm}
\centerline{\psfig{file=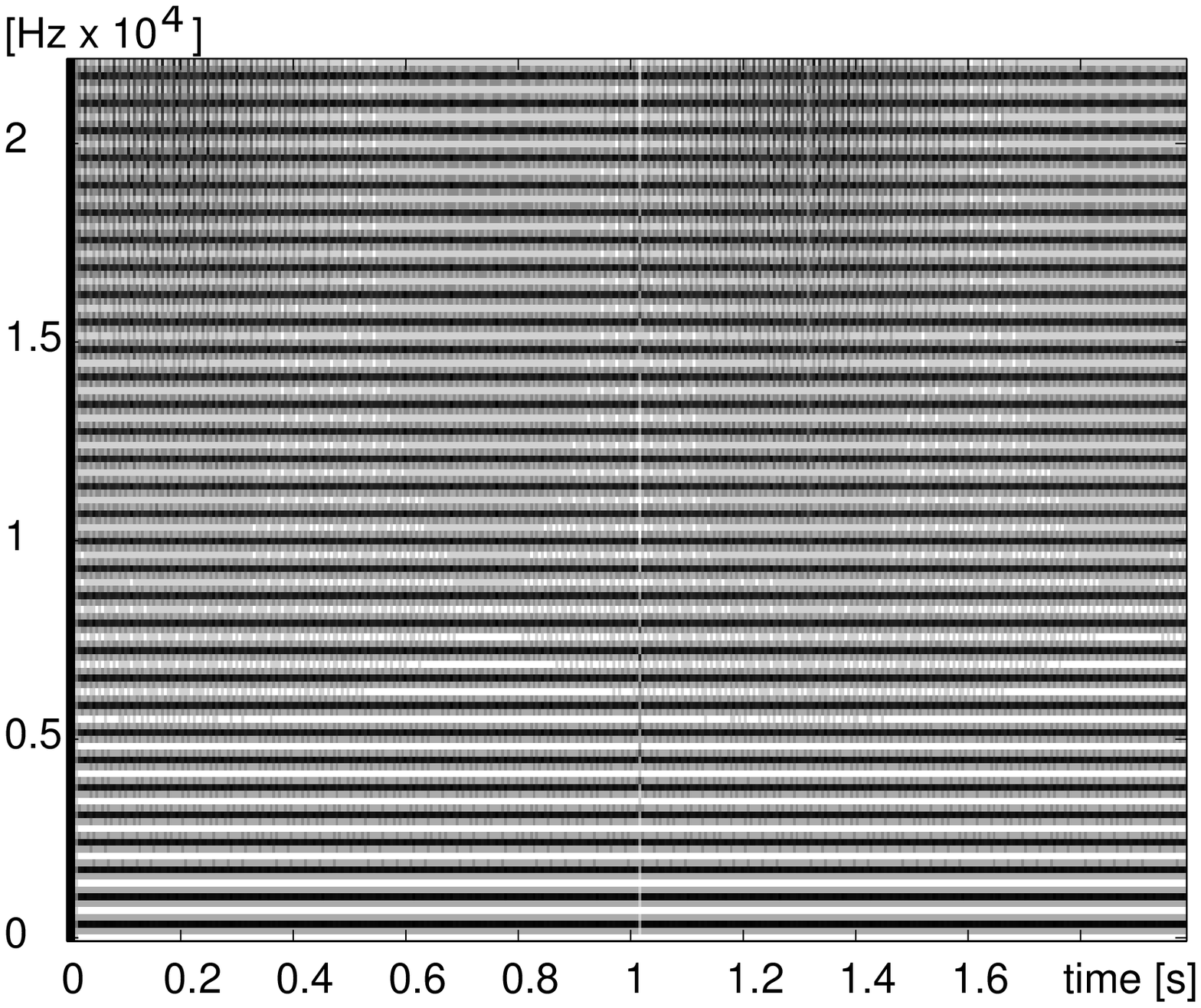,width=8.0cm}}
\centerline{(b)}
\caption{Sonogram of the  response of a non-modulated (a) and modulated (b) FIR delay line with linearly-increasing length to a pulse train. Hanning windows of 256 samples are used in analysis. Magnitude (in dB) is smaller where the points are darker.}
\label{noisyramp}
\end{figure}
\fi

A different issue is the audibility of the side components introduced by modulation. This depends on the strength of the side components and on their position relative to the carrier signal. If these components lay below the threshold of masking for the carrier signal~\cite{ZwickerFastl} they are not audible. If the modulation frequency is around $60-80 Hz$ the side components turn out to be below that threshold for partials laying in the first $2-3 \hbox{kHz}$. 

\section{A Fractionally-Addressed Delay Line}
\label{fadline}
An alternative realization of the delay line can be developed by observing that a single pointer is sufficient for  both the read and write accesses. If the delay line has fixed integer
length $B$, it is possible to use a buffer exactly $B$-cells long and
a single pointer whose entry is first read and then written. At every sample the phase pointer is incremented to point to the following cell. In the
same buffer we can also implement any delay which is an integer
fraction $B/I$ just by incrementing the pointer at steps of $I$
samples.
We are going to show how this scheme can be generalized to non-integer
fractions of the total buffer length. The resulting technique can be
seen as an extension of the table-lookup oscillator~\cite{Moore77,Hartmann87},
with the fundamental difference that every read is followed by one or more writes, in such a way that
the waveform is continuously re-stored while being read.

Given a buffer size of $B$ samples, and a sample rate $F_s$, a
(fractional) increment of $I$ samples gives a delay in seconds equal
to
\be
T = \frac{B}{I \cdot F_s} \, .
\label{delsec}
\ee 

Since this realization is related to waveform generation by fractional
addressing~\cite{Hartmann87}, we call it the Fractionally-Addressed
Delay (FAD) line.

\subsection{Frame-Based Realization}
The most immediate way to code the FAD line is by means of frame-based processing, e.g., using the {\tt resample} function of the MATLAB Signal Processing Toolbox\footnote{MATLAB is a registered trademark of The MathWorks, Inc., http://www.mathworks.com}. The code chunk of figure~\ref{framecode} is intended to be inserted into the framework of a frame-based digital audio effect, as prescribed by the COST-G6 action on Digital Audio Effects (DAFX)\footnote{See the COST-G6 DAFX home page \\ http://echo.gaps.ssr.upm.es/COSTG6/}~\cite{Arfib98}. It uses a {\tt buffer} of length {\tt lenbuf} to implement a delay having fractional length {\tt Q/P * lenbuf} (i.e. $I = \hbox{\tt lenbuf}/\hbox{\tt delay} = {\tt P/Q}$).
\if F\draft
\begin{figure}
{\tt
\begin{verbatim}

delay = Q/P * lenbuf; %delay in samples
framelen = floor(delay);
for n=1:nframes
  bufout = resample(buffer, P, Q)'; %read
  output = [[output, bufout]];
  fwrite(fid_out, bufout, 'int16');
  bufin = fread(fid_in, framelen, 'int16');
  buffer = resample(bufin, Q, P); %write
end
\end{verbatim}
}
\caption{MATLAB code for a frame-based realization of the FAD line}
\label{framecode}
\end{figure}
\fi

At every frame, the {\tt buffer} is first {\bf read} with a decimating factor equal to {\tt P/Q} and written to the output file {\tt fid\_out}. Then {\tt framelen} values of the input stream, coming from file {\tt fid\_in}, are interpolated with factor {\tt P/Q} and {\bf written} into the {\tt buffer}. The {\tt resample} operation, which uses a polyphase implementation of the interpolation filters~\cite[pages 677--679]{Mitra}, takes care of changing the effective length of the buffer to a fractional number, in such a way that the output stream is the {\tt delay}-ed version of the input stream. According to equation~(\ref{delsec}), using an increment ranging from $2 $ to $1$ we can implement any fractional delay ranging from $B/2$ to $B$ samples, where $B$ is the buffer length. For the purpose of this paper we limit the variability of the delay length to half its nominal size. Nothing prevents to reduce the length even further, but this is not computationally convenient in the implementation presented in section~\ref{samplebysample}.

The implementation of figure~\ref{framecode} shows the close relationship between sample-rate conversion~\cite{Mitra,Orfanidis,Zolzer} and delay interpolation. This connection can be useful to understand the fractional delay thoroughly. For instance, we usually adopt $I>1$ because otherwise we would write a downsampled (and therefore bandlimited) version of the signal into the buffer, thus making it impossible to recover the high-frequency components in the following read.
\Comment{\it Ricordarsi nelle modulazioni di suggerire incrementi altini per spostare in alto i prodotti di modulazione.}
 
In the implementation of figure~\ref{framecode}, continuous variations of the effective delay length can not be imposed within a single frame, since the interpolation factor is varied frame by frame. Continuous delay modulations are essential for constructing digital audio effects such as choruses and flangers~\cite{Bloom}, so that a frame-by-frame MATLAB implementation of these effects should use alternative interpolating functions where the interpolation factor can be prescribed in a local per-sample basis (see e.g. the MATLAB function {\tt interp1}).
However, there are many applications of fractional delay lines within signal processing flow-graphs exhibiting feedback-connected modules. In general, these modules can not be computed on a frame-by-frame basis due to circular dependencies spanning time delays different from the frame length. Therefore, we are mainly interested in devising a sample-by-sample implementation of the FAD line and to study its behavior in terms of signal accuracy and computational complexity.
  
\subsection{Sample-by-Sample Realization}
\label{samplebysample}
As we have mentioned, the FAD line can be interpreted as an extension of the table-lookup oscillator, where the waveform is written while it is being read. The read operation can be treated in exactly the same way as in the table-lookup oscillator, being possible to apply truncation, polynomial interpolation, or multirate
interpolation techniques~\cite{Orfanidis,Zolzer}.
More complicated is the injection of a new value, to be done right
after the read, in such a way that no ``holes'' are left in the current
pass through the buffer. A fractional increment would correspond to a variable number of writes at each step. For instance, for $I=1.5$, three writes have to
be performed for every couple of reads. 
Interpolation in write has been proposed by V\"alim\"aki et al. in the context of digital waveguide modeling~\cite{VesaFilters,VesaT}, and there called deinterpolation. \Comment{A deinterpolator is in fact the transposed time-reversed form of an interpolator, and is implemented in FIR form as a sequence of cumulative additions  into a delay line.} A deinterpolator can be obtained from a FIR interpolator by transposing its structure and inverting the sequence of coefficients~\cite[pages 128--134]{VesaT}. These operations lead to a structure which is a sequence of cumulative additions into a delay line. Equivalently, interpolation can be performed on the input stream by using some extra unit delays, thus saving read and write operations in the buffer. This latter implementation is depicted in figure~\ref{buffer} for the case of second-order read and write interpolators, with an increment $I = 1.25$. The dash-dotted lines in the figure should indicate how resampling is performed on the fly on both the read and the write side of the access to the buffer. If further writes have to be performed after the first one in order to fill the blanks, the quantity {\tt d} of figure~\ref{buffer} has to be updated to {\tt d + 1} and the corresponding coefficients of the writing interpolator have to be recomputed.

\if F\draft
\lungo{
\begin{figure}[htb]
\psfig{file=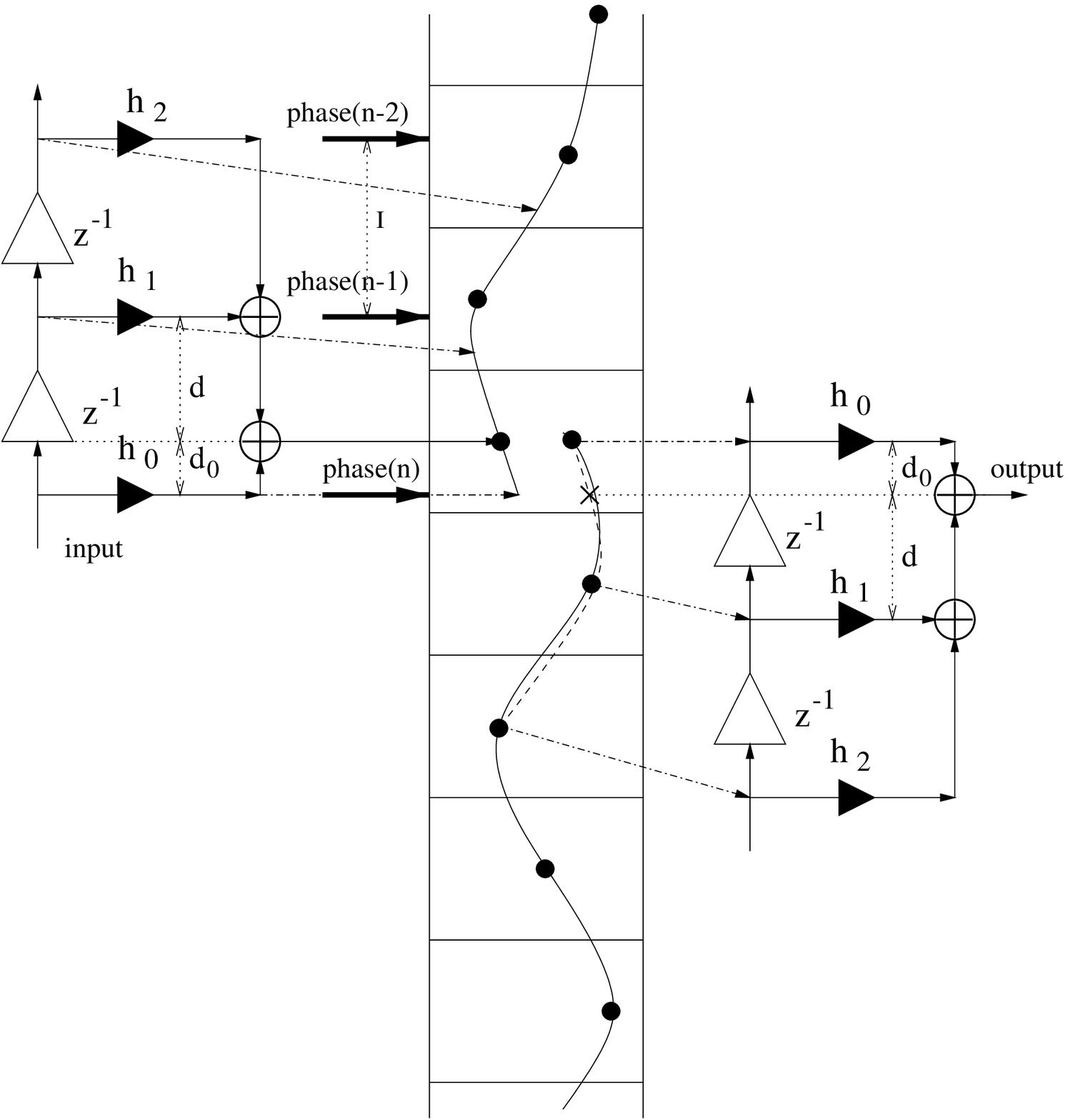,width=8.0cm}
\caption{Interpolated read and write access to a circular buffer}
\label{buffer}
\end{figure}
}
\fi

\Comment{Several interpolation
techniques can also be applied at the write stage.}

The FAD line can be expressed by the pseudo-code of figure~\ref{samplecode}. Actual MATLAB and C functions implementing the FAD line can be found in the software repository of the  COST-G6 action on Digital Audio Effects\footnote{The software repository can be reached from the home page of the COST-G6 DAFX action: \\http://echo.gaps.ssr.upm.es/COSTG6/}.

\if F\draft
\begin{figure}
\begin{tt}
\begin{verbatim}
loop
     fph = floor(phase);
     output = interpolated_read(table[fph],
             table[fph+1], ...);
     ph = (phase_old + 1) MOD length_table;
     while (ph <= fph) {
       table[ph] = interpolated_write(
         ..., table[phase_old], input);
       ph = (ph + 1) MOD length_table;
     }
     phase_old = fph;
     phase = (phase + Increment);
     if (phase > length_table)
          phase = phase - length_table;
endloop
\end{verbatim}
\end{tt}
\caption{Pseudo-code for a sample-by-sample realization of the FAD line}
\label{samplecode}
\end{figure}
\fi

Notice that the {\tt interpolated\_read} uses samples following the {\tt phase} pointer, while the {\tt interpolated\_write} uses samples preceding the pointer.

In the following sections we are going to analyze the performance of a FAD line using quadratic interpolation in both read and write operations. \Comment{, as compared to the classic implementation of a delay line with second-order Lagrange interpolation.}

\section{Input-Output Analysis}
\label{io}
\subsection{Experimental results}
The FAD line is a time-varying system, and therefore it is difficult
to characterize in terms of frequency response. \Comment{However, it is worth
noticing how it responds to a sinusoidal input.} \lungo{Figure~\ref{outi} shows
the magnitude spectrum of a delayed $5-{kHz}$ sine wave when quadratic interpolation
is used in reading and writing.} When a sinusoidal input feeds the FAD line, \Comment{
It is clear that} spurious components are added to the main spectral
line\Comment{~\cite{roccim98}}. The magnitude of these components might be dependent on the
frequency of the input sine wave and the initial (fractional) phase of the FAD-line
pointer~\footnote{As an example of dependence on initial phase, consider the increment $I = 1$. If
the initial phase is $0$ the pointer always falls on samples. If the
initial phase is $0.5$ the pointer always falls between samples. In the two cases, the actual shape of the delayed output is different.}. The signal-to-error noise ratio (SNR) as a function of these two
parameters shows a very mild dependence on initial phase. Therefore,
it makes sense to plot the average SNR as a function of the input
frequency only (figure~\ref{SNRcompared}). We can see that low
frequencies are affected by high SNR, thus indicating that the FAD
line has an acceptable behavior for practical
sounds. 
\Comment{Figure~\ref{SNRcompared}
also shows a comparison with the
quadratically-interpolated (FIR) delay line.} The noise error has been computed as the sum of the
squared differences between the input and output waveform samples\Comment{ by considering four full cycles of the sinewave}\footnote{A normalizing factor $\sqrt{2/N}$ has been applied, being $N$ the number of samples per period.}~\cite{Moore77}. In figure~\ref{SNRcompared} we have considered input sine waves having periods that perfectly divide the delay length, and a time delay that is two thirds of the delay length (increment is $1.5$). By applying such SNR analysis to a low-order FIR filter we would obtain a curve lower than that of figure~\ref{SNRcompared}~\cite{RocchessoDAFX98}. We do not report that curve here because it can be misleading. In fact, while the error of the FAD line can be interpreted as noise, the error for a FIR line is only given by phase displacement and magnitude attenuation, and no spurious signal components are introduced.
\Comment{The noise error is larger in the FIR case even though that implementation has no
spurious components in the output spectrum. This is due to the fact that the FIR line exhibits a larger phase distortion for that particular fractional delay.} \Comment{This is due to the fact
that  the FIR attenuation is larger on average.}
\Comment{More generally, the contribution of the nonlinearity of phase should also be taken into account. }
\Comment{In practice, the actual SNR curve would stay all below the drawn line.}

\if F\draft
\lungo{
\begin{figure}[htb]
\psfig{file=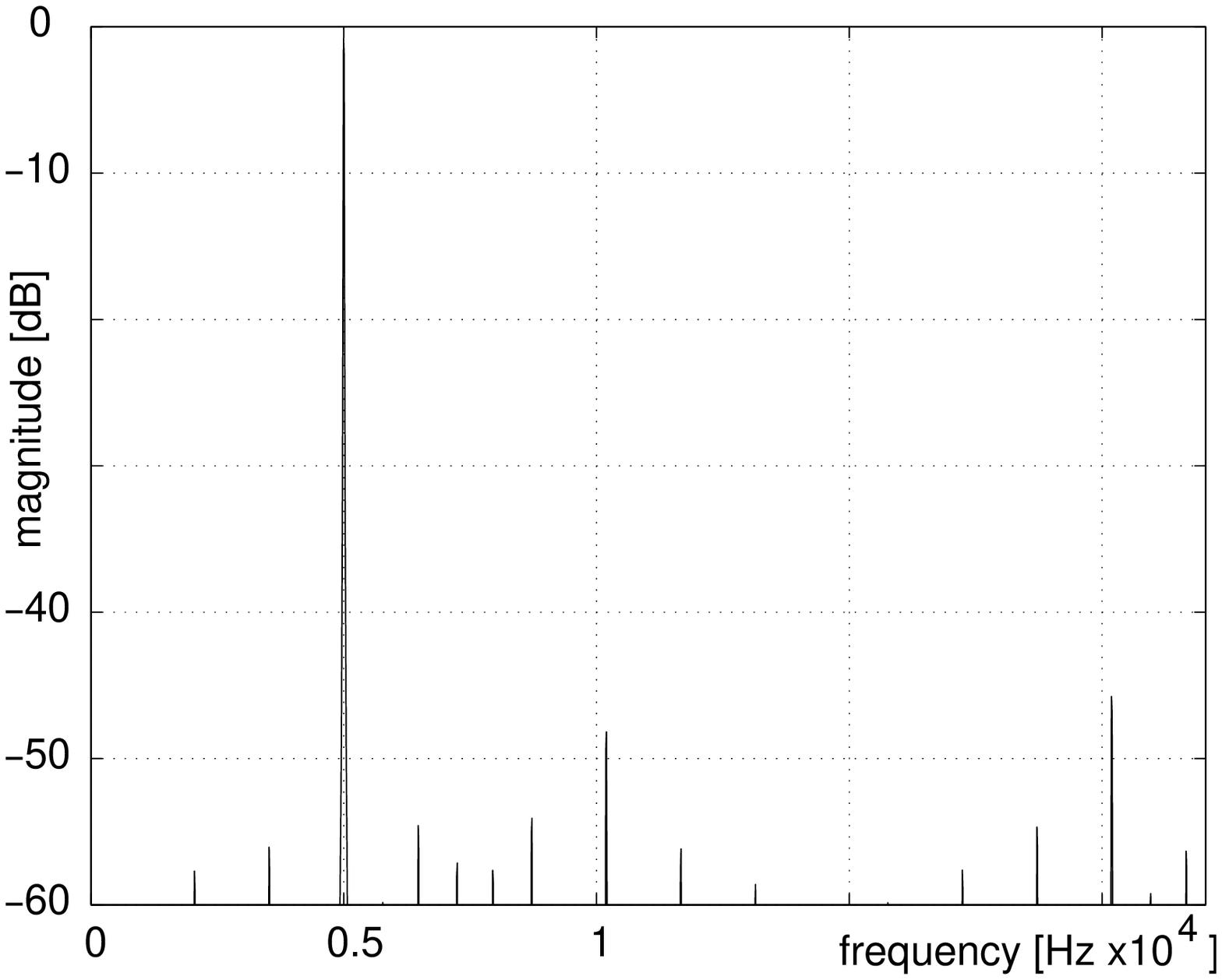,width=8.0cm}
\caption{Magnitude spectrum of the output signal of a FAD line, where the
input signal is a sine wave at $5000 \hbox{Hz}$, the delay is $0.74378
\hbox{s}$, the sampling rate is $44.1 \hbox{kHz}$ and the buffer is $44100-\hbox{samples}$ long.  }
\label{outi}
\end{figure}
}
\fi

\Comment{\begin{figure}[htb]
\epsfysize=9.0cm
\epsfxsize=7.0cm
\centerline{\hfill\rotate[r]{\mbox{{\epsfbox{SNR.ps}}}}\hfill}
\caption{Signal-to-error noise ratio as a function of initial phase and sine frequency}
\label{SNR}
\end{figure}
}

\if F\draft
\begin{figure}[htb]
\psfig{file=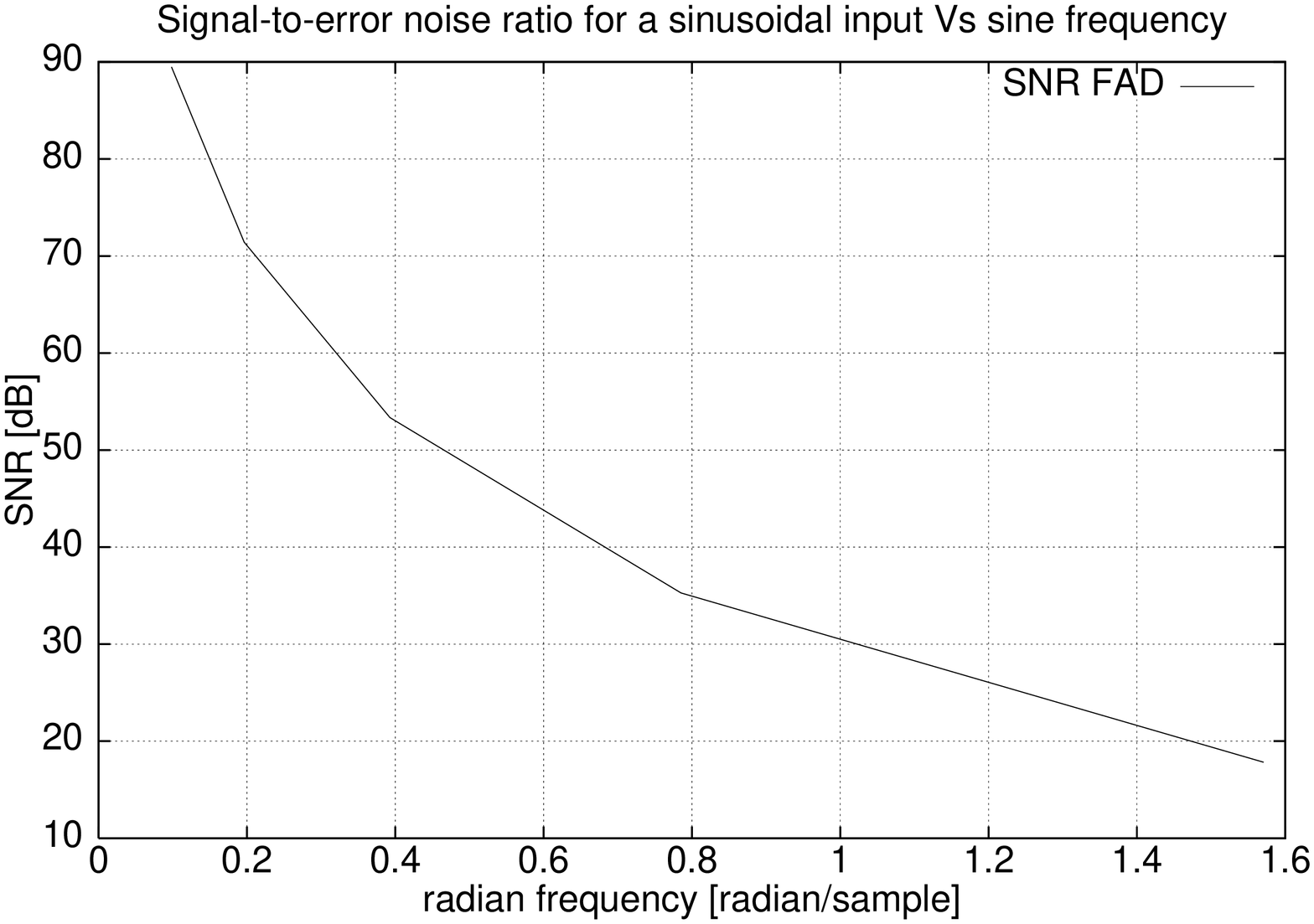,width=8.0cm,angle=-90}
\caption{Experimental signal-to-error noise ratio Vs. sine frequency for the
FAD line  with quadratic interpolation. 
The input frequencies are such that the interpolation phase is a multiple of $2\pi$.}
\label{SNRcompared}
\end{figure}
\fi

\Comment{It should be noticed that the spurious components tend to cluster around the main peak, thus being likely to be below the masking threshold, whose slope is known to be less than $27 \hbox{dB} / \hbox{Bark}$~\cite{ZwickerFastl}.}

Especially for applications such as waveguide modeling of musical
instruments~\cite{SmithInBrandenburg98}, it is important to  consider 
the  attenuation that different
frequencies are subject to when fed into the delay line. For instance, the decay time of a frequency partial in a waveguide string depends on the attenuation of the interpolated delay line at that frequency.
The attenuation of the main peak of the output spectrum turns out to
be dependent on the initial phase. In order to have a rough idea of the attenuation property of the quadratically-interpolated FAD line, we plot \Comment{compare it with the linearly-interpolated FIR line by plotting}  in figure~\ref{attcompared} the
minimum, maximum, and mean attenuation as a function of frequency
of the input sine wave, for the same set of frequencies as in figure~\ref{SNRcompared}, and for a delay such that $I = 1.5$.
\Comment{Notice that the linearly-interpolated FIR line shows an attenuation that is significantly higher than that  of 
the quadratic FAD line.} Again, direct comparison with the FIR line is problematic because the results would be highly dependent on the chosen delay length.

\Comment{A more detailed plot showing the attenuation for high-pitch sinewaves is reported in figure~\ref{attcomparedhf}, which has been computed experimentally on eight cycles of the input and output sinewaves. }

\Comment{\begin{figure}[htb]
\epsfysize=9.0cm
\epsfxsize=7.0cm
\centerline{\hfill\rotate[r]{\mbox{{\epsfbox{att.ps}}}}\hfill}
\caption{Attenuation of a sinusoidal input as a function of initial phase and sine frequency}
\label{att}
\end{figure}
}

\if F\draft
\begin{figure}[hbt]
\psfig{file=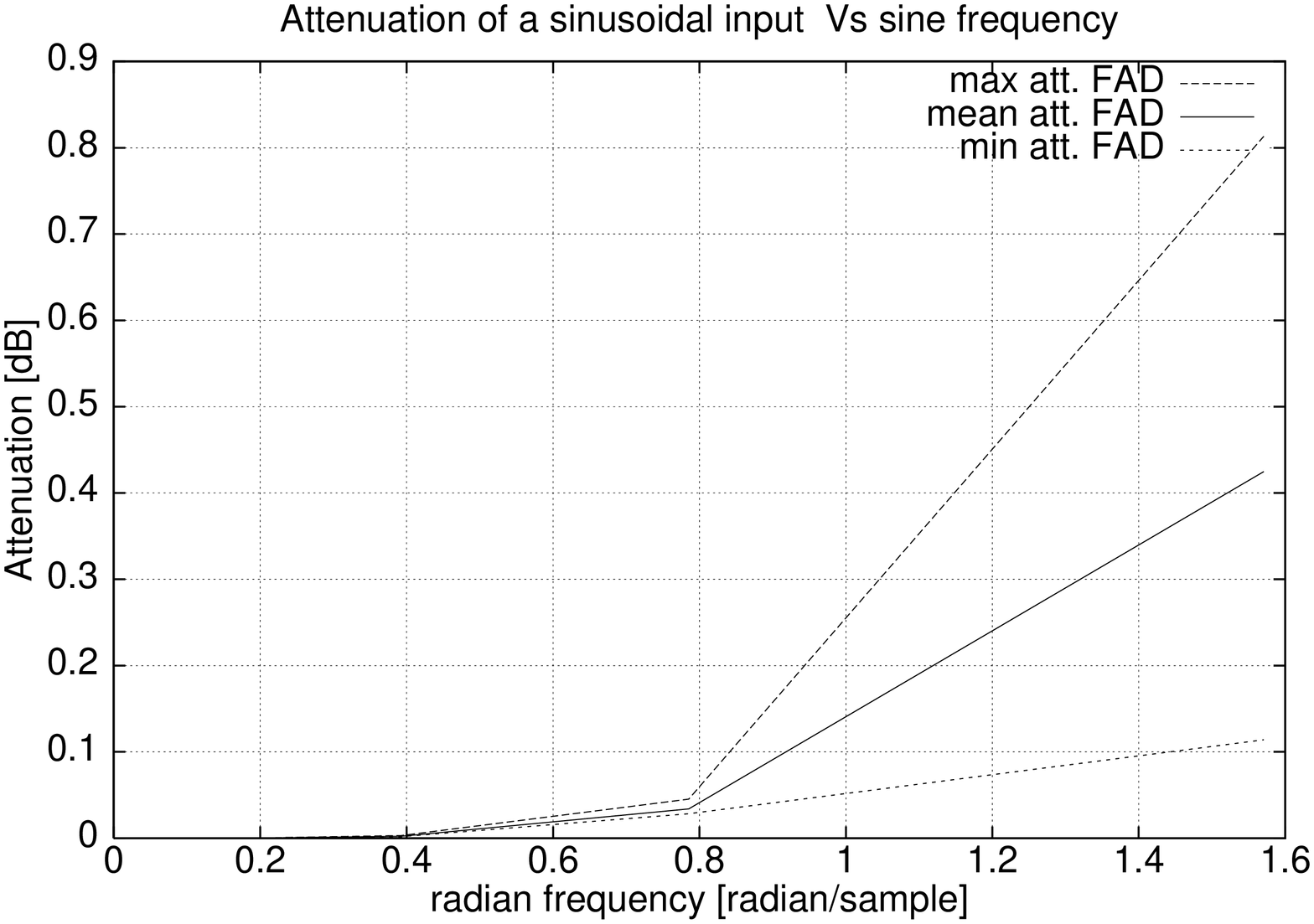,width=8.0cm,angle=-90}
\caption{Attenuation of a sinusoidal input Vs. sine frequency for the
quadratically-interpolated FAD line. Sampling rate is set equal to the buffer length and delay is set to $D = 2/3 \hbox{s}$}
\label{attcompared}
\end{figure}
\fi

\Comment{
\begin{figure}[hbt]
\epsfysize=8.0cm
\epsfxsize=4.0cm
\centerline{\hfill\rotate[r]{\mbox{{\epsfbox{snrzoom.ps}}}}\hfill}
\caption{Attenuation of a sinusoidal input Vs. sine frequency for the
FIR line and the FAD line}
\label{attcomparedhf}
\end{figure}
}

\Comment{
\begin{figure}[ht]
\epsfysize=9.0cm
\epsfxsize=7.0cm
\centerline{\hfill\rotate[r]{\mbox{{\epsfbox{outni.ps}}}}\hfill}
\caption{Spettro di uscita, sinusoide a 8000 Hz, ritardo di 0.74831
sec, interpol. quadratica in scrittura e nessuna interpol. in lettura}
\label{outni}
\end{figure}

\begin{figure}[ht]
\psfig{file=outils.ps,width=8.0cm,angle=-90}
\caption{Spettro di uscita, sinusoide a 8000 Hz, ritardo di 0.74831
sec, interpol. lineare in scrittura e lineare in lettura}
\label{outils}
\end{figure}
}

\subsection{Signal-to-error noise analysis}
\Comment{A Lagrange interpolator can be characterized at any fractional delay by its magnitude and phase delay responses~\cite{LaaksoB}. At any frequency $f$ and for a given interval of fractional delays, the interpolator has a magnitude response ranging from $A_{\rm max}$ and $A_{\rm min}$ and a phase delay ranging from  $\tau_{\rm max}$ and $tau_{\rm min}$. For example, a second-order (quadratic) interpolator can be constructed by using the coefficients~\cite{VesaT}:
\beqa
h_0 & = & d (1 + d) /2 \nonumber \\
h_1 & = & (1 + d) (1 - d) \\
h_2 & = & -d (1 - d) /2 \nonumber
\eeqa 
}
In section~\ref{lengthmod} we have seen how delay-length modulations add side components to the peaks of a frequency spectrum, as a result of modulations in magnitude and phase responses of the Lagrange interpolator.
In the implementation of the FAD line, the magnitude and phase delay of the interpolators are varied sample by sample, as the value of $d$ (see figure~\ref{buffer}) is changed at every sample. If we exclude the degenerate cases (such as that obtained with increment $I=1$) we can assume that, in consecutive accesses to the buffer, the magnitude and the phase delay vary sinusoidally around their mean. 
Therefore, we can use the approximate analysis of section~\ref{lengthmod}, where an input  sinewave at frequency $\omega_0$ is subject to phase modulation with a certain modulating frequency $\omega_M$ and to amplitude  modulation with frequency $2 \omega_M$.

The modulating frequency $\omega_M$ is related to the fractional part $I_f$ of the increment $I$ by 
\be
\omega_M = 2 \pi I_f F_s \, .
\ee
If $I_f \stackrel{>}{\sim} 0.3$, the modulation products are folded into the high-frequency region for practical sounds, i.e. for sounds whose energy is mainly concentrated in a bandwidth smaller than $F_s/8$. This means that these modulation products are less audible (especially when the sample rate is larger than $44 kHz$), and it is easier to eliminate them by lowpass filtering. 

\Comment{
The amplitude-modulated signal can be expressed at every frequency as
\beqa
v_{AM} & = & \left(A_m + \frac{1 - A_{\rm min}}{2} \cos{\omega_M t}\right) \cos{\omega_0 t} \nonumber\\
& = & A_m \left(1 + m \cos{\omega_M t}\right)\cos{\omega_0 t} \nonumber\\
& = & A_m  \cos{\omega_0 t} + \\
& & + \frac{m A_m}{2} \cos{(\omega_0 - \omega_M t} +  \frac{m A_m}{2} \cos{(\omega_0 + \omega_M t)} \nonumber
\label{AM}
\eeqa
where $A_m = (A_{\rm min} + A_{\rm max})/2$ and $m$ is the modulation index 
\be
m = \frac{1 - A_{\rm min}}{1 + A_{\rm min}}
\ee
The equation~(\ref{AM}) shows that in the output spectrum there is a main peak and two modulation products, as it was found in figure~\ref{outi}. 
}

\Comment{
The signal-to-error noise ratio can be found as the ratio between the RMS values of the modulation products and the value of the main peak, and it assumes the value
\be
SNR_{AM} = \frac{\sqrt{2}}{m}
\ee
In the FAD line there are two interpolations acting in cascade. The input signal is affected by two amplitude modulations having the same modulating frequency. In the worst case, the two modulations cooperate constructively by further multiplying the main peak by $A_m$ and by doubling the modulation products. If the effect on SNR of other modulation products is neglected the new signal-to-error noise ratio is
\be
SNR_{AM_FAD} = \frac{A_m}{m\sqrt{2}}
\ee
}
\Comment{Figure~\ref{snrAM} depicts these signal-to-error noise ratios as a function of the input frequency.
\begin{figure}[hbt]
\epsfysize=8.0cm
\epsfxsize=4.0cm
\centerline{\hfill\rotate[r]{\mbox{{\epsfbox{snrAM.ps}}}}\hfill}
\caption{Signal-to-error noise ratio for the quadratic Lagrange interpolator and for the quadratically-interpolated FAD line}
\label{snrAM}
\end{figure}
}
\Comment{The phase-modulated signal can be expressed at every frequency $\omega_0$ as
\beqa
v_{PM} & = & \cos{\left(\omega_0 t + \tau_{\rm max}\omega_0 \sin{\omega_M t}\right)} \nonumber\\
& = & J_0 \cos{\omega_0 t} + \nonumber \\
&  & - J_1 \cos{(\omega_0 - \omega_M t} + J_1 \cos{(\omega_0 + \omega_M t)} \nonumber \\
&  &  + J_2 \cos{(\omega_0 - 2 \omega_M t} + J_2 \cos{(\omega_0 + 2 \omega_M t)} \nonumber \\
&  &  - J_3 \cos{(\omega_0 - 3 \omega_M t} + J_3 \cos{(\omega_0 + 3 \omega_M t)}
\label{PM}
\eeqa
where $J_i$ is the $i$-th order Bessel function of the first kind evaluated in $ \tau_{\rm max}\omega_0$.
}

As opposed to the case analyzed in section~\ref{lengthmod}, in the FAD line there are two modulations, one acting on input and the other acting on output. However, we can still use the values of table~\ref{modprod} to compute an estimate of the SNR of the FAD line, in the pessimistic case when the two modulations produced by the read and write interpolators operate constructively. An approximate value of this SNR is
\be
SNR  = \frac{{A_0}^2}{2 \sqrt{2} \sqrt{{A_1}^2 + {A_2}^2 + \dots }} \; ,
\ee
\Comment{
\be
SNR_{PM} = \frac{J_0}{\sqrt{2} \left(J_1 + J_2 + J_3 + \dots \right)}
\ee
}
where the carrier appears at the numerator and the sidebands sum up at the denominator.
Figure~\ref{snrPM} depicts this signal-to-error noise ratio as a function of the input frequency. The curve gives a tight lower bound to the  values of SNR measured in simulations and reported in figure~\ref{SNRcompared}, thus justifying the assumptions taken in order to simplify the analysis.
\if F\draft
\begin{figure}[hbt]
\psfig{file=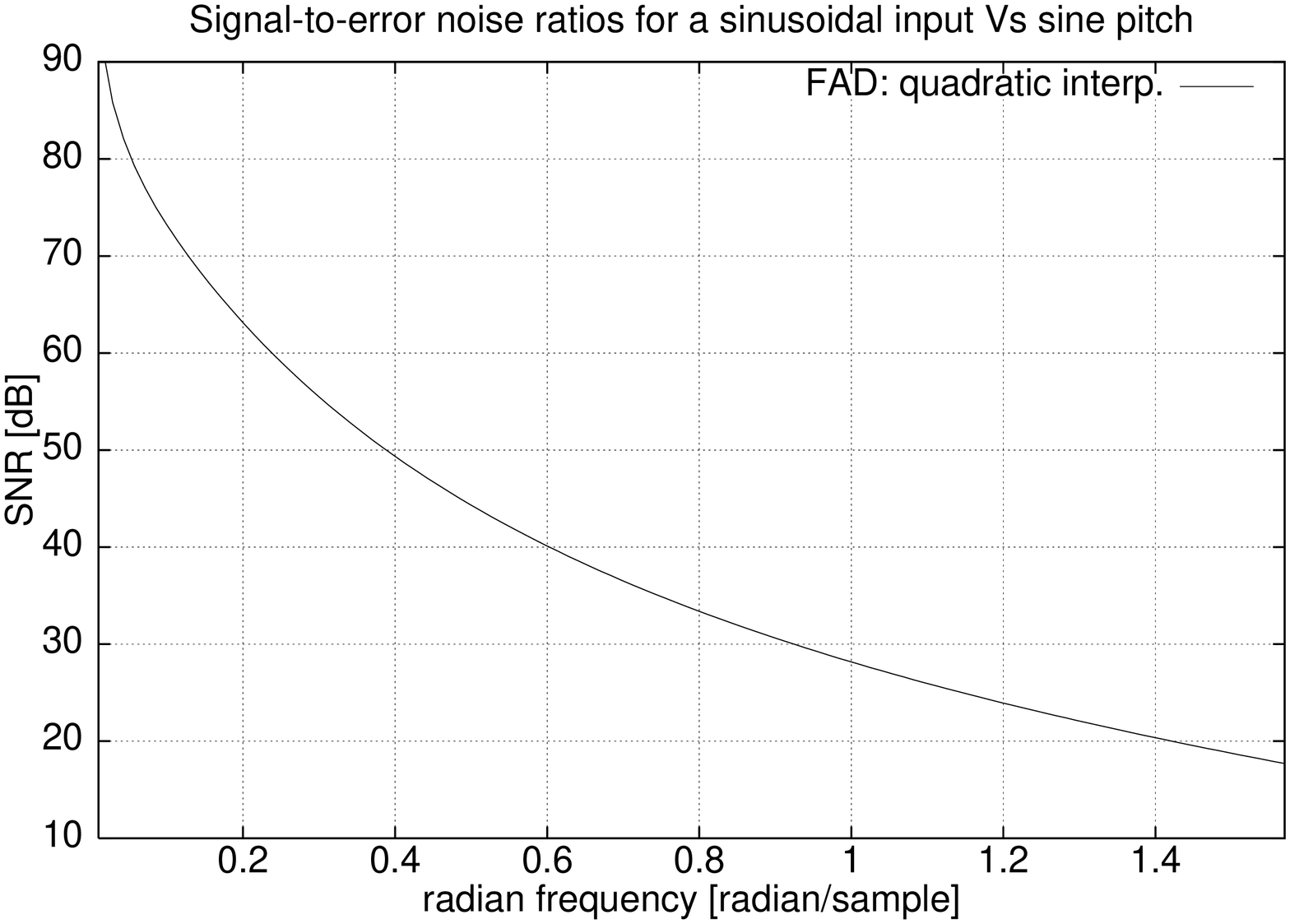,width=8.0cm,angle=-90}
\caption{Signal-to-error noise ratio for the quadratic FAD line }
\label{snrPM}
\end{figure}
\fi

\subsection{Behavior for time-varying delay}
\label{timevar}
The FAD line shows an unconventional behavior when the delay length is
dynamically varied\Comment{~\cite{roccim98}}. This is illustrated in the following analysis through comparison with the FIR implementation.

Starting from the steady state of a delay line fed with a stationary signal, suppose to vary the delay length as a
linear function of time $t$. Namely, we start at time $0$ with the nominal delay $\tau_0$
and decrease it at the rate of $k$ seconds per second:
\be
T(t) = \tau_0 - k t \; .
\label{rampa}
\ee
The FIR implementation responds with an instantaneous pitch shift in the output
signal. In other words, we get a Doppler effect and the pitch shift is 
\be
\Delta f = 1 + k \; .
\ee
On the other hand, the FAD line provides a steady pitch shift
\be
\Delta f = e^k 
\label{deltaf}
\ee
after a transient time  
\be
\tau_i = \frac{\tau_0}{k}(1 - e^{-k}) \; .
\label{ttime}
\ee
A similar transient is observed when the delay ramp is stopped.

The transient time~(\ref{ttime}) can be calculated by feeding the delay line with an
impulse at time $0$. It will come out of the line at time instant
$\tau_i$ such that
\Comment{
\be
\int_0^{\tau_i} \frac{1}{T(t)} dt = 1
\ee
}
\be
\int_0^{\tau_i} I(t) dt =  \frac{B}{F_s} \; , 
\label{int1}
\ee
where $I(t)$ is the time-dependent increment which produces the desired ramp in delay length. Equation~(\ref{int1}) can be rewritten, using~(\ref{delsec}) and~(\ref{rampa}), as
\be
\int_0^{\tau_i} \frac{1}{\tau_0 - k t} dt = 1 \; ,
\ee
which is solved by~(\ref{ttime}).

The steady-state transposition~(\ref{deltaf}) can be calculated by observing that a second
impulse entering the line at time $T_i$ ``sees'' an instantaneous  delay of $\tau_0 - k
T_i$ seconds. It gets out of the line at time $\frac{\tau_0 - k
T_i}{k}(1 - e^{-k}) + T_i$, exactly $T_i e^{-k}$ seconds after the
impulse which entered at time 0. 

It is interesting to notice that the dynamic behavior of the FAD line is similar to the behavior of the analog CCD delay line~\cite{Catrysse80}, which was built in MOS technology as a sampling system where packets of charge were shifted through the channel of a multigate MOSFET transistor\footnote{Thanks to Giovanni De~Poli for pointing out this similarity.}. In that case, if $B$ was the number of gates and $F_s$ was the control frequency of a multiphase clock, the delay was still given by~(\ref{delsec}), provided that $I$ is set to one. In order to vary the delay of the CCD line, $F_s$ had to be made time varying, and it is easy to see that~(\ref{deltaf}) and~(\ref{ttime}) held for that line as well.

\lungo{If the delay ramp is applied for 1.11 seconds starting at time $0$ with an empty line,
the response of the FAD line to a steady sinusoidal input is displayed
as a sonogram in figure~\ref{gliss}. The transient is clearly visible in the output
when the ramp is stopped.  This sonogram should be compared to the one
obtained with the FIR line (figure~\ref{glissf}).

\if F\draft
\begin{figure}[ht]
\psfig{file=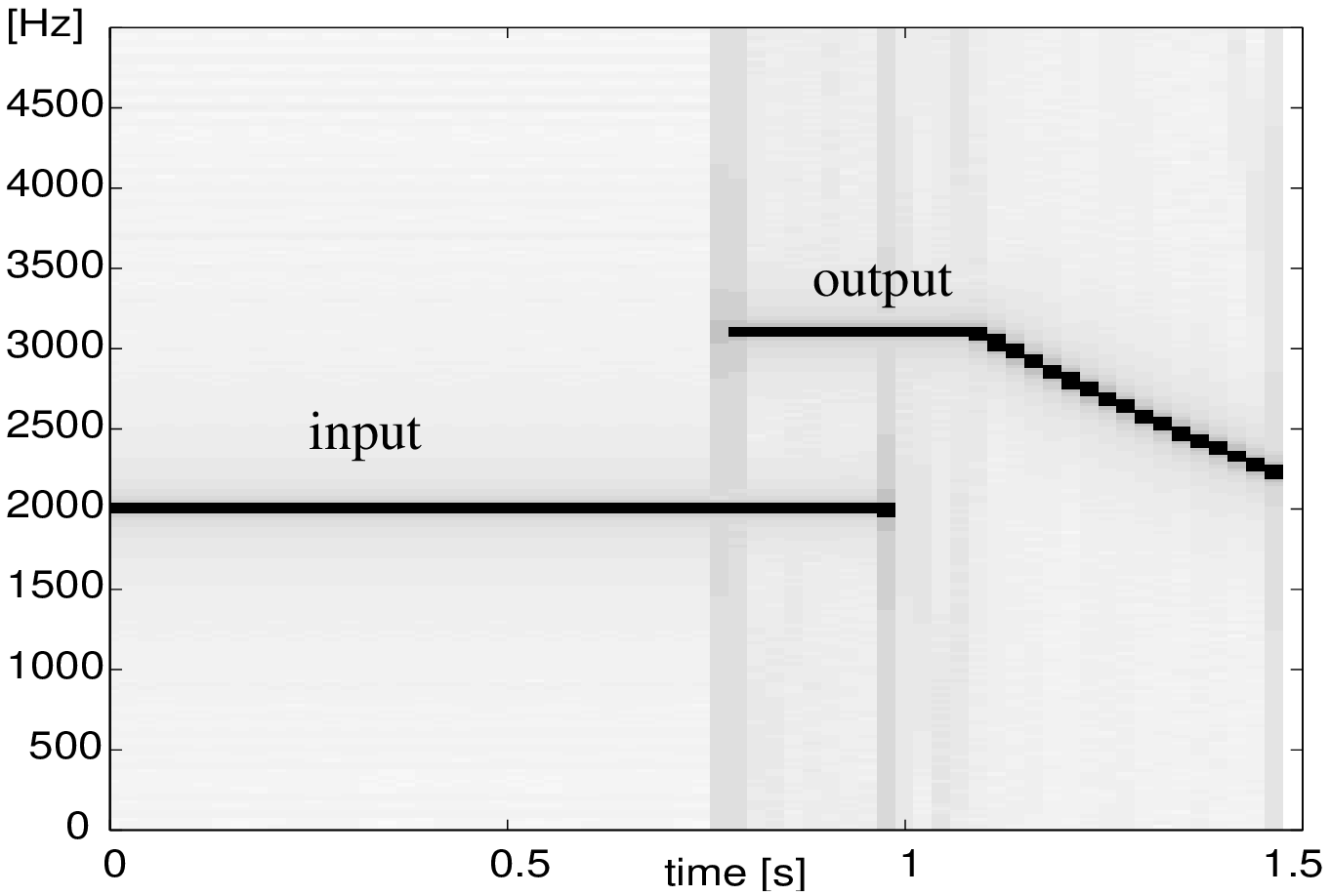,width=8.0cm}
\caption{FAD line: delay ramp from 0.99 s to 0.5 s in 1.11 s; 1 s of sinusoidal input}
\label{gliss}
\end{figure}

\begin{figure}[ht]
\psfig{file=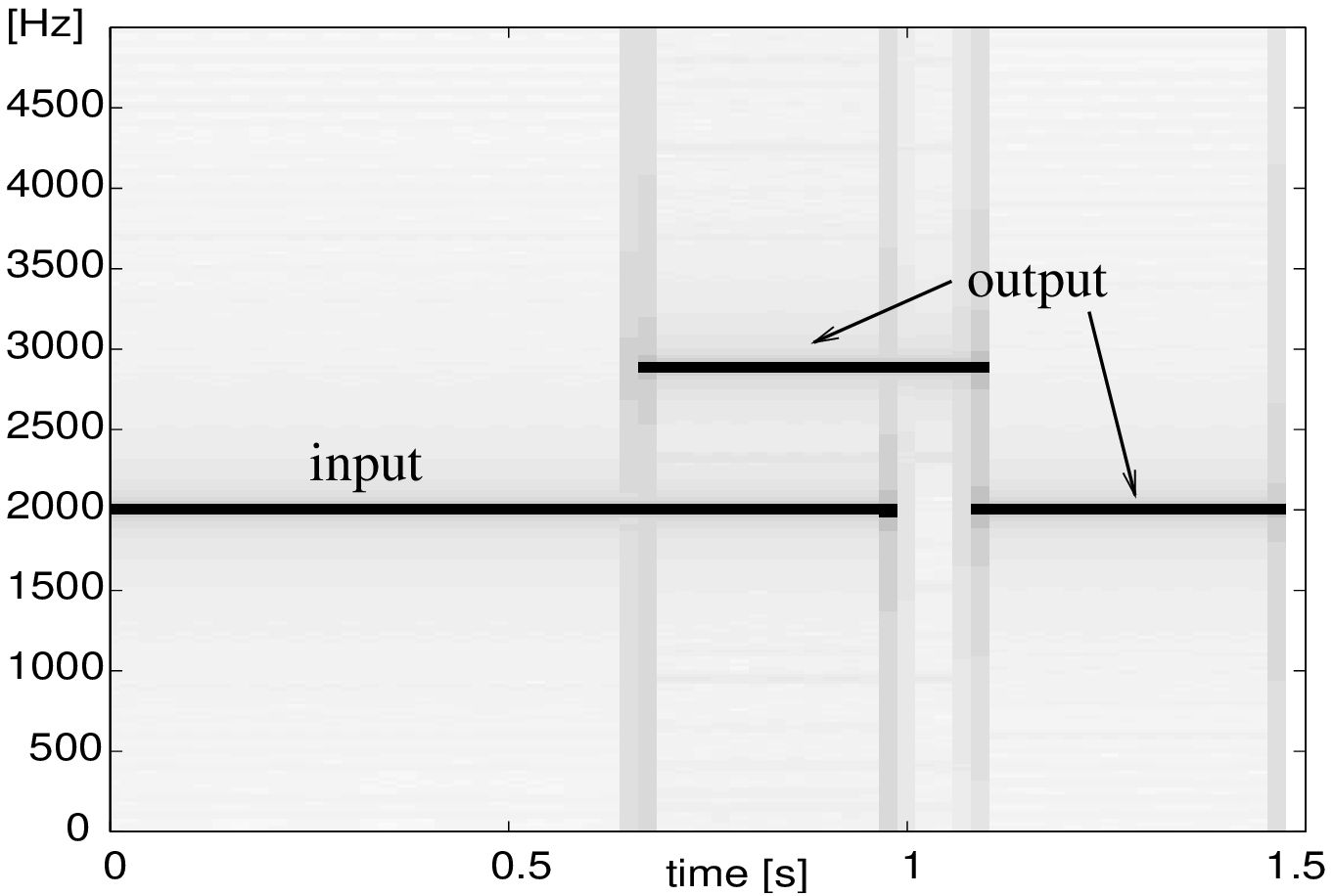,width=8.0cm}
\caption{FIR line: delay ramp from 0.99 s to 0.5 s in 1.11 s; 1 s of
sinusoidal input} 
\label{glissf}
\end{figure}
\fi
}

A different behavior is also reported in response to sinusoidal
modulations of the delay length\Comment{~\cite{roccim98}}. These modulations are essential for
effects such as flanging or phasing. \lungo{Modulations of the quadratic FAD and FIR
lines are reported in figure~\ref{vibrfa} and~\ref{vibrfir},
respectively. The figures show that the FAD line is less
sensitive to artifacts clearly visible (and audible) as faint waves 
figure~\ref{vibrfir}, which are essentially due to modulations
induced by the nonideal response of the interpolator.

\if F\draft
\begin{figure}[ht]
\psfig{file=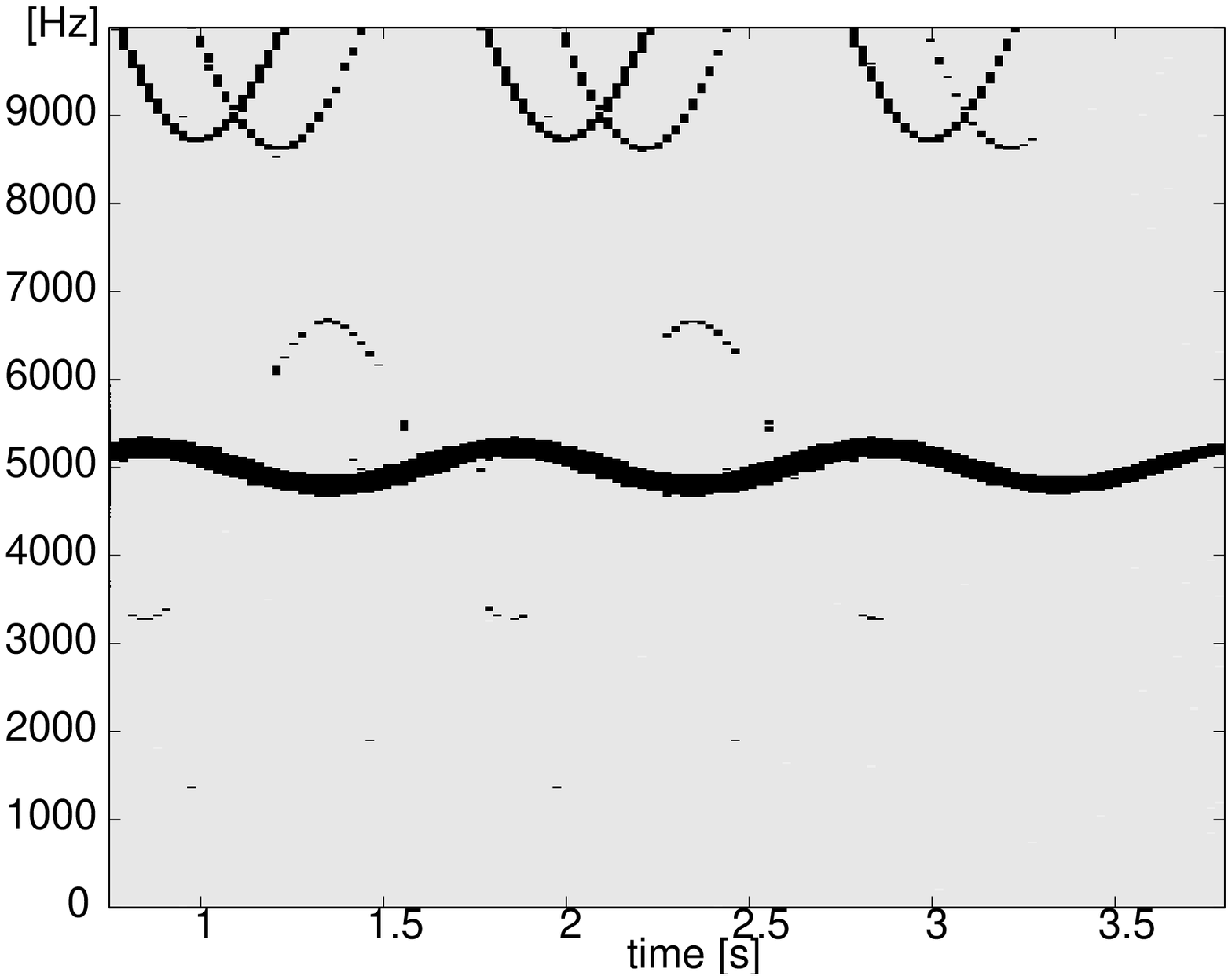,width=8.0cm}
\caption{FAD line with quadratic interpolation: delay-length vibrato}
\label{vibrfa}
\end{figure}

\begin{figure}[ht]
\psfig{file=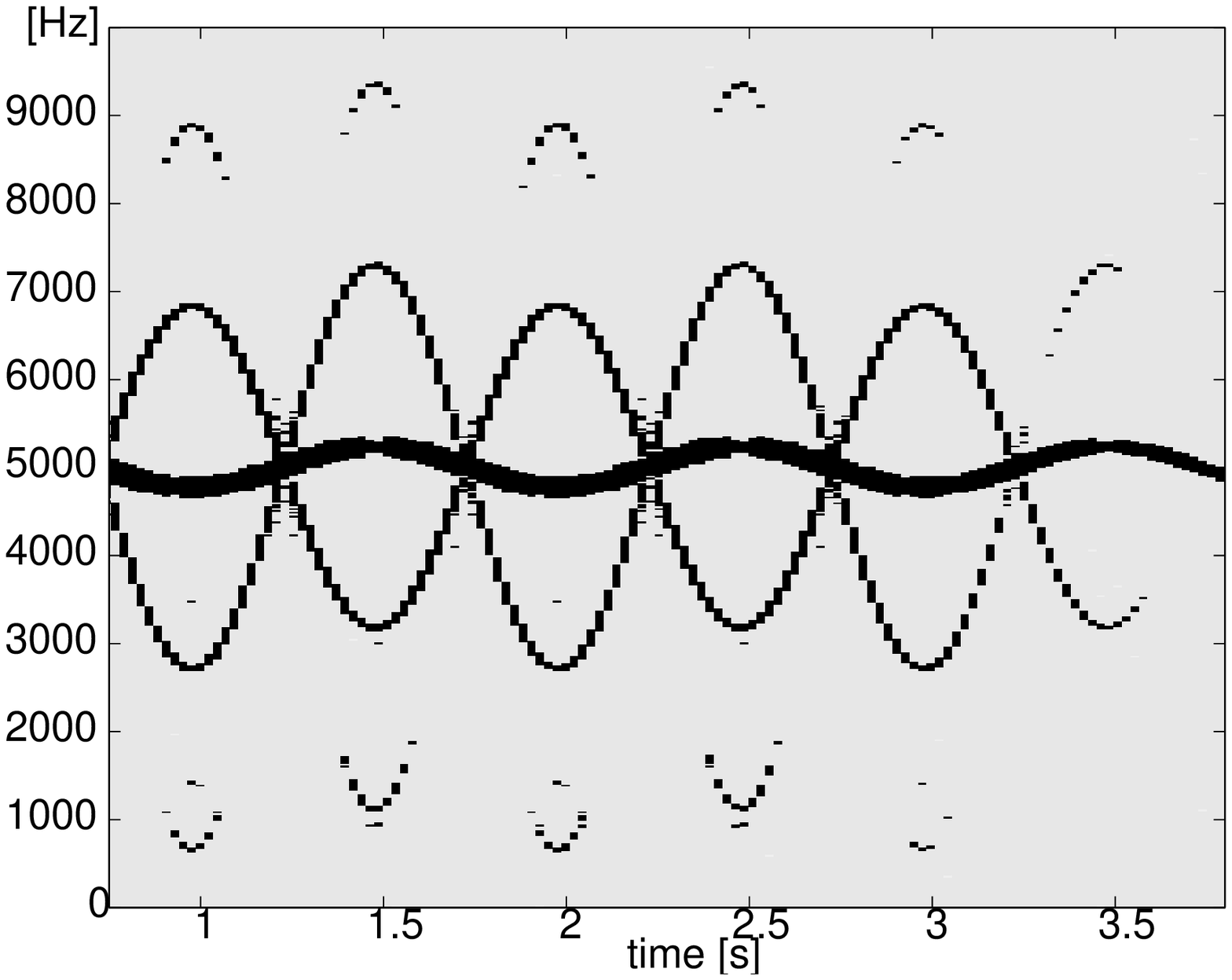,width=8.0cm}
\caption{FIR line with quadratic interpolation: delay-length vibrato} 
\label{vibrfir}
\end{figure}
\fi
}

\section{Physical Interpretation}
\label{physical}
If the dynamic behavior of the delay lines is closely analyzed, we see that
the FAD and the FIR realizations actually simulate two different
physical phenomena. In both cases, the lines can be thought of as a
one-dimensional medium where waves propagate. However, when the delay
length is dynamically reduced we have two different physical analogies in the
two cases. The shortening of the FIR line corresponds to the receiver
getting closer to the transmitter, and therefore we have a tight
simulation of the Doppler effect. On the other hand, the shortening of the FAD line
corresponds to increasing the speed of propagation in the medium
while maintaining the same physical distance between the two
ends. \Comment{This is analogous to what happens if we increase the tension of
a guitar string by turning the tuning pin. }

Figures~\ref{waveguidestring}--\ref{ntfad} illustrate what happens when using different implementations of the delay lines in the simulation of a one-dimensional waveguide resonator, such as a string. In this application, sketched in fig.~\ref{waveguidestring}, there is a couple of delay lines in a feedback connection, each representing propagation of waves in one direction. For the sake of simplicity, the terminations are supposed to be perfectly reflecting. If one of the delays is fed by  three periods of a fast sine wave, this packet propagates, gets reflected, feeds the other line, and comes back to the excitation point for another reflection. Under ideal conditions, the packet keeps going back and forth without attenuation or losses. Suppose that, right after a reflection at the left end, suddenly the string gets lengthened by some amount, as we would do for lowering the pitch of the string. This can be simulated, in the classic FIR line implementation, by moving the reading pointer backwards. However, this operation exposes again the wave packet which has just been reflected, thus modifying the ``duty cycle'' of the waveform, as reported on figure~\ref{ntfirne}. \Comment{The correct waveform would be that of figure~\ref{ntfire}, which} A correct waveform can be obtained by adding a write operation right after the read to the FIR line implementation. This write takes care of erasing the waveform as it passes the reading pointer. 
Another way of lowering the pitch is that of changing the string tension instantaneously. In a waveguide simulation this corresponds to changing the spatial sampling via a change in the waveguide speed of propagation~\cite{SmithInBrandenburg98}. This can be achieved by the FAD line implementation just by changing the phase increment, and the result is illustrated in figure~\ref{ntfad}.
A mixture of tension and length increase might be obtained by using the FAD line with a variable buffer size.

A different approach to dynamic tension variations in string models has been recently proposed by V\"alim\"aki et al.~\cite{ValimakiICMC98,ValimakiICASSP99}. They recast the resampling process induced by tension modulation into delay length modulation of a FIR line controlled by the buffer content. This reformulation works under the assumption of a single observation point. The control circuitry is composed of a power estimator (which needs to sum the squares of the content of several delay cells) and a numerical integrator. 

\if F\draft
\begin{figure}[htb]
\psfig{file=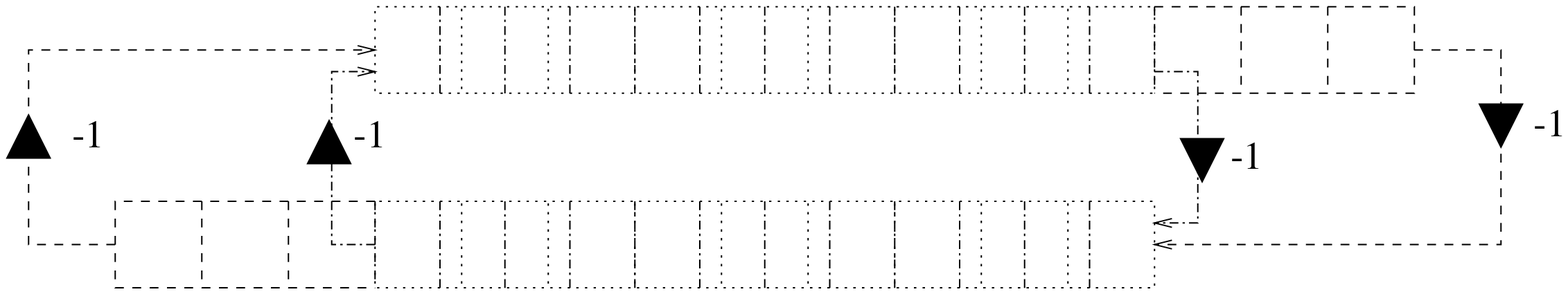,width=8.0cm}
\caption{Waveguide model of an ideal string. A sudden pitch lowering is depicted for the FIR implementation of delay lines (dashed line) and for the FAD implementation of delay lines (dash-dotted line). The string before the pitch transition is represented by the dotted line.}
\label{waveguidestring}
\end{figure}

\begin{figure}[hb]
\psfig{file=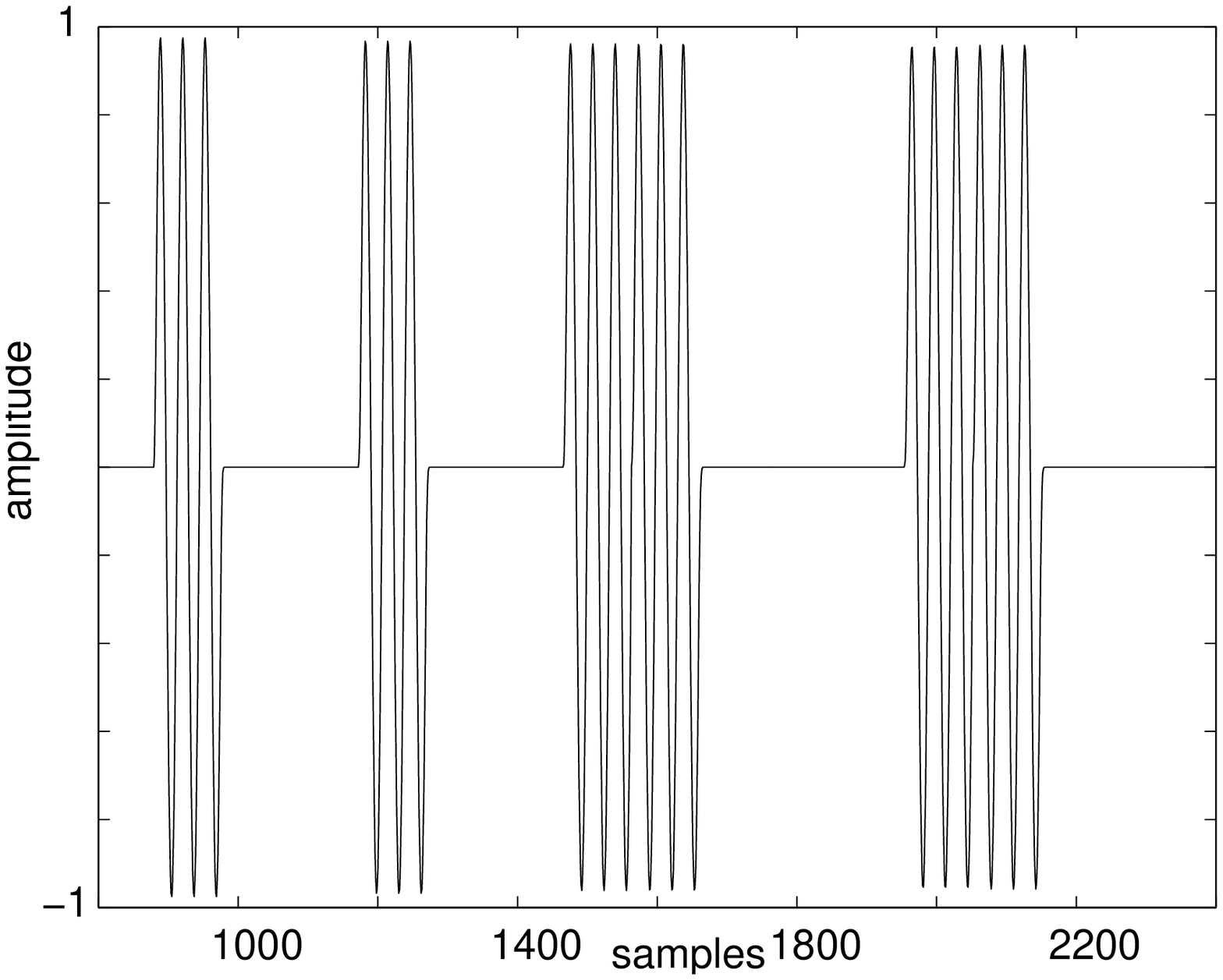,width=8.0cm,height=4.0cm}
\caption{Note transition waveform: FIR line without erase after read}
\label{ntfirne}
\end{figure}

\begin{figure}[ht]
\psfig{file=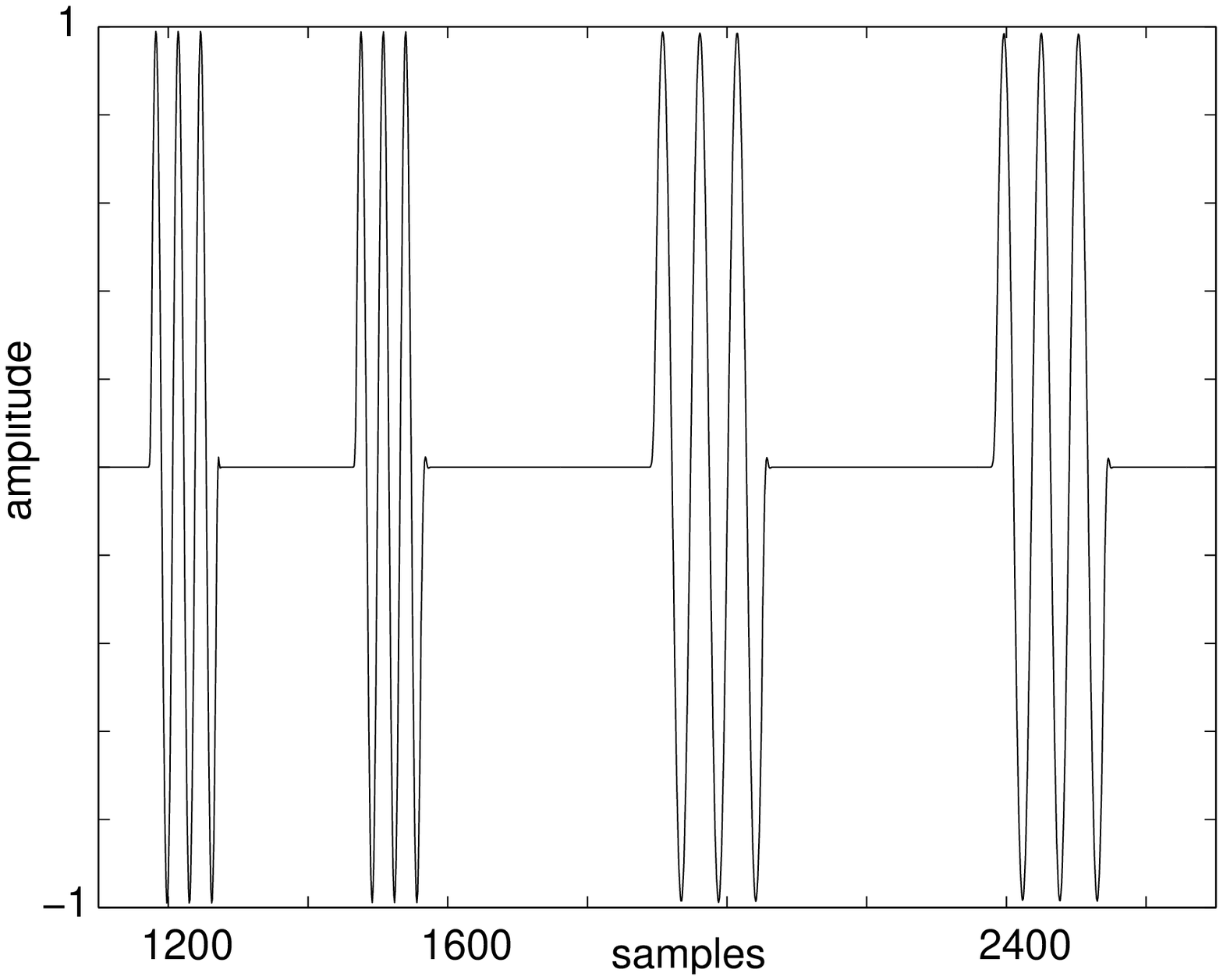,width=8.0cm,height=4.0cm}
\caption{Note transition waveform: FAD line}
\label{ntfad}
\end{figure}

\Comment{\lungo{\begin{figure}[hb]
\psfig{file=pitch3.wav.ps,width=8.0cm,angle=-90}
\caption{Note transition waveform: FIR line with erase after read}
\label{ntfire}
\end{figure}
}
}
\fi

The two ways of producing a pitch shift are not equivalent as far as timbre is concerned, as it can be deduced from figures~\ref{ntfirne} and~\ref{ntfad}. In fact, \Comment{\lungo{it can be seen from the sonograms in figures~\ref{ntfadson}--\ref{ntfireson} that}} the FAD-line pitch change produces a contraction of the whole spectrum, thus modifying the position of formants, while the FIR-line pitch change comes without moving the formants. \Comment{\lungo{In figure~\ref{ntfirneson} it is also visible the artifact produced},} Moreover, when no waveform erasing is applied out of the reading pointer in the FIR line,  the width of the main formant gets narrowed. On the other hand\Comment{, figure~\ref{ntfireson} shows that} erasing after read preserves both the formant widths and positions.

Summarizing, different implementations of the delay line do have practical consequences on the timbre produced by dynamically-varying waveguide models. 

\if F\draft
\lungo{
\Comment{
\begin{figure}[htb]
\psfig{file=pitch1.son.ps,width=8.0cm}
\caption{Note transition sonogram: FAD line}
\label{ntfadson}
\end{figure}

\begin{figure}[htb]
\psfig{file=pitch2.son.ps,width=8.0cm}
\caption{Note transition sonogram: FIR line without erase after read}
\label{ntfirneson}
\end{figure}

\begin{figure}[htb]
\psfig{file=pitch3.son.ps,width=8.0cm}
\caption{Note transition sonogram: FIR line with erase after read}
\label{ntfireson}
\end{figure}
}
}
\fi

\section{Computational Performance and Application to Digital Audio Effects}
\label{performance}
The FAD line has only one pointer for accessing data in the buffer. It
exhibits spatial locality because  any short sequence of accesses spans over
a small neighborhood of the pointed buffer cell. On the other hand, a FIR
line has two pointers, thus exhibiting two distinct spatial localities.
As a consequence, we expect that the FAD line makes better use of the cache in
general purpose computer architectures. 
However, the FAD performs more writes than reads. In order to attain a
$50\%$ of delay variability, we have to accept up to two writes for
each read. \lungo{This overhead is partially compensated by the highest efficiency of write operations in modern
architectures~\cite{Duvanenko98}.}

\lungo{These two observations justify the fact that} the FAD line, despite of
its higher complexity, does not run much slower than the FIR line on a
general purpose computer. A benchmark for quadratically interpolated FIR and FAD lines has been performed on an
$AMD-K6$ architecture by repeatedly delaying a soundfile stored in an
array. The experiment was done using a Linux operating system with the machine in stand-alone single-user configuration. \Comment{The experimental conditions are realistic because several processes were active without loading the machine, thus giving the context switches that one is likely to find in practice.} \Comment{The first run was neglected because it seemingly involves instruction and data loading in the cache.} To avoid the effects of context switching due to the underlying operating system, we took the fastest of 14 repetitions, thus obtaining the  results 
summarized in figure~\ref{benchk6}\Comment{table~\ref{bench}}, where running times are reported for varying buffer size. \Comment{We see that the performance for the FAD 
line is even better than that of the FIR line if a simple interpolation scheme is used.}

A first comment is about the  difference in performance between the two algorithms. This is not as big as one might expect, especially if we consider that the quadratically-interpolated FAD line has about three times as many multiplies, twice as many adds, three extra divides, and three times as many tests as the FIR line.
A second comment is about the fact that the curves tend to be monotonically increasing. This indicates that more and more cache misses are encountered when using larger buffers. However, the FIR line shows an increase in capacity and conflict misses right after the size  of 65536 samples which, when translated in bytes\lungo{ (we use 8-byte double floating point numbers)}, gives the size of the level-2 cache\footnote{And also of the memory covered by the Translation Lookaside Buffer, which is responsible for fast translation of virtual addresses to physical addresses.}. 

\Comment{The fact that the slope of the FIR curve is less regular and locally higher than the other curve confirms that having only one locality helps\footnote{However, this phenomenon doesn't show up when the same benchmark runs on an Intel Pentium II.}. In the FIR curve it is also possible to see two peaks of steepness right before reaching the sizes of 4096 and 65536 samples which, when translated in bytes\lungo{ (we use 8-byte double floating point numbers)}, give respectively the size of the level-1 data cache (32 KBytes) and the size of the level-2 cache\footnote{And also of the memory covered by the Translation Lookaside Buffer, which is responsible for fast translation of virtual addresses to physical addresses.}. }
Using the FIR line and eliminating the phase increment it is possible to measure the overall cost of caching, wich is  visible from figure~\ref{benchk6} as the difference between the two lower curves. Since it turns out to be less than 6\% even for very large buffers, we can argue that delay lines are not affected much by the memory hierarchy.
Similar values of the percentage cost of caching have been measured in other architectures (e.g., the Intel Pentium II), even though the actual shape of the curves is slightly different. 

In our implementations, we have not done aggressive code optimization, so that the relative performance of the three realizations might vary in practice from what we have shown. In particular, float-to-integer conversions are expensive, and it would be wise to perform them by direct bit manipulation, as suggested in~\cite{Dannenberg97d}. Moreover, the memory access patterns show that hardware or software prefetching techniques might be used effectively, especially in the FAD line that has only one locality.

\Comment{
In any case, our results show that alternative implementations might be considered for sound-processing building blocks, even when the cost in terms of pure floating point operations seems daunting. 
}
\if F\draft
\begin{figure}[htb]
\psfig{file=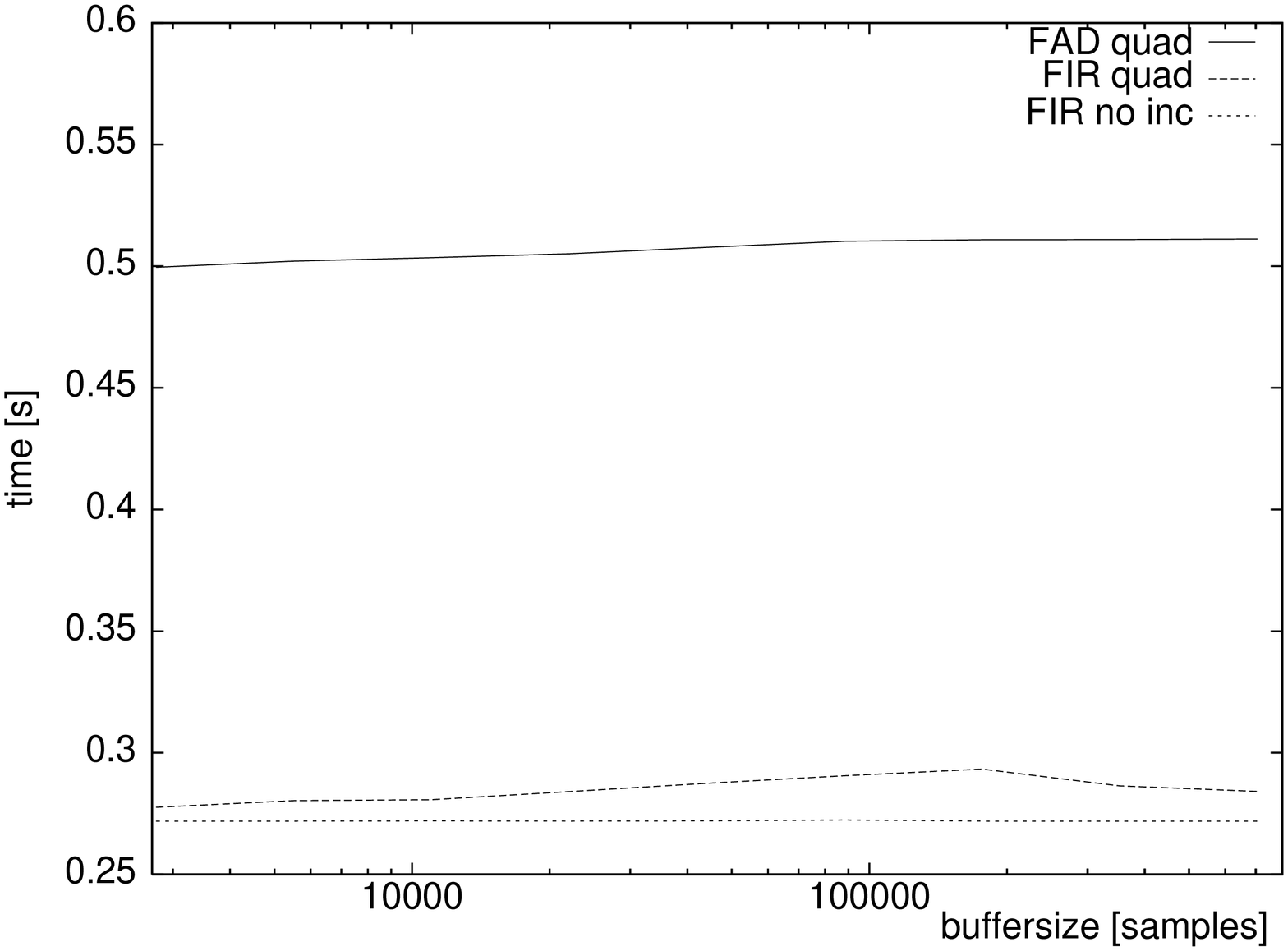,width=8.0cm,angle=-90}
\caption{Performance on an AMD-K6 of the quadratically-interpolated FAD line, FIR line, and  FIR line with no phase increment, as a function of buffer size.}
\label{benchk6}
\end{figure}
\fi

\Comment{
\begin{figure}[htb]
\epsfysize=8.0cm
\epsfxsize=5.0cm
\centerline{\hfill\rotate[r]{\mbox{{\epsfbox{p2.ps}}}}\hfill}
\caption{Performance on an Intel Pentium-II of the quadratically-interpolated FAD line, linearly-interpolated FAD line, linearly-interpolated FIR line, and FIR line with no phase increment, as a function of buffer size.}
\label{benchp2}
\end{figure}
}

\Comment{ 
\begin{table}
\begin{tabular}{|p{5.7cm}|r|}
\hline
\multicolumn{2}{|c|}{Benchmark}
\\ \hline \hline
Delay Line & time (s) \\
\hline \hline
FAD line, linear interpolation in write and read & 1.00 \\
FAD line, quadratic interp. in write and linear interp. in read & 1.21
\\
FIR line, linear interpolation & 1.14 \\
\hline
\end{tabular}
\caption{Benchmark for different implementations of the delay line on
a general-purpose computer}
\label{bench}
\end{table}
}

High-quality digital audio effects, such as choruses or flangers, can be easily constructed around delay-modulated FAD lines. The audio quality is good due to  good noise rejection for time-varying delay, as shown in sec.~\ref{timevar}. 
The fact that the FAD line does not respond instantaneously to variations of the delay length turns out to be useful for achieving natural sounding effects by means of random variations of the delay length. For instance, ``random walk'' pitch modulations can easily be achieved by low-rate random variations of the delay length. Namely, it is sufficient to call a random function, say, every 100 samples, and to perform  piecewise linear interpolation between its values. The resulting output spectrum exhibits partials which are floating around without sudden changes. In the FIR implementation, a similar effect would require either a complicated modulating function to be read at audio rate or 
high-order  interpolation  performed on the undersampled control signal~\cite{FernandezDAFX98}.

\roc{Fare i conti su come va la trasposizione per cambiamenti bruschi di delay.

Fai il confronto con lavoro di Pablo.}

\section{Conclusion}
We have proposed a realization of the digital delay line which is
based on an extension of the table-lookup oscillator. The proposed
realization exploits the features of modern computer architectures and
shows good performance in terms of signal-to-error noise ratio, frequency-dependent attenuation
and dynamic behavior. We expect this delay line will be
considered as a building block for physically-based sound synthesis
and for sound effects such as flangers and choruses.

\bibliography{../../general}

\if T\draft

\clearpage

\section*{Tables}
\begin{table}[h]
\begin{tabular}{|c|c|c|} \hline
$A_0$ & $A_1$ & $A_2$ \\ \hline\hline 
& & \\
$A_m J_0 + {m A_m} J_2 $ & $ A_m J_1 - \displaystyle{\frac{m A_m}{2} J_1} $ & $ \displaystyle{A_m J_2 + \frac{m A_m}{2} J_0} $ \\ 
& & \\ \hline
\end{tabular}
\caption{Amplitude of the sidebands of an amplitude- and phase-modulated signal}
\label{modprod}
\end{table}

\section*{Illustrations}
\clearpage
\begin{figure}[ht]
\centerline{\psfig{file=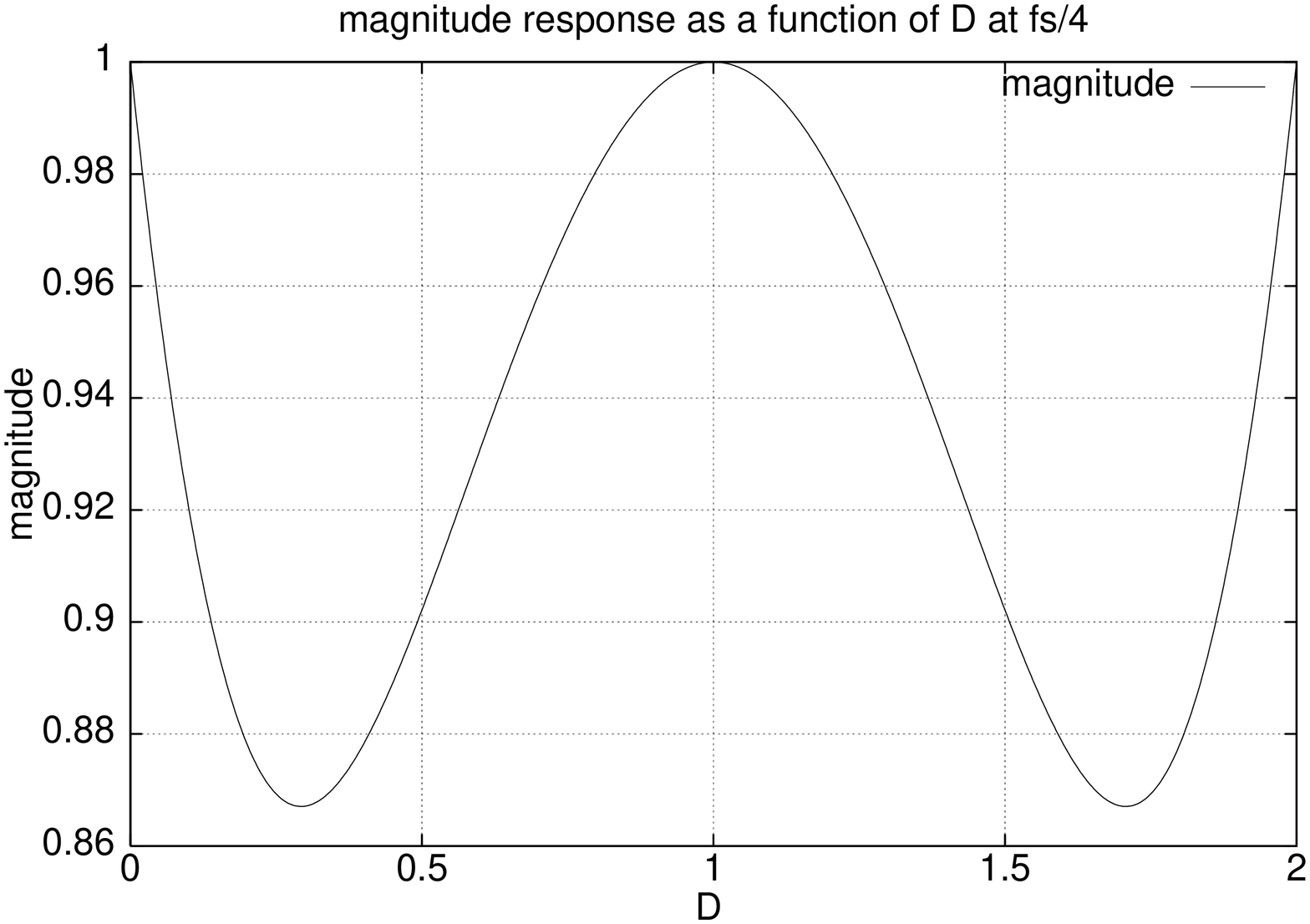,width=11.0cm,angle=-90}}
\centerline{(a)}
\vspace{0.1cm}
\centerline{\psfig{file=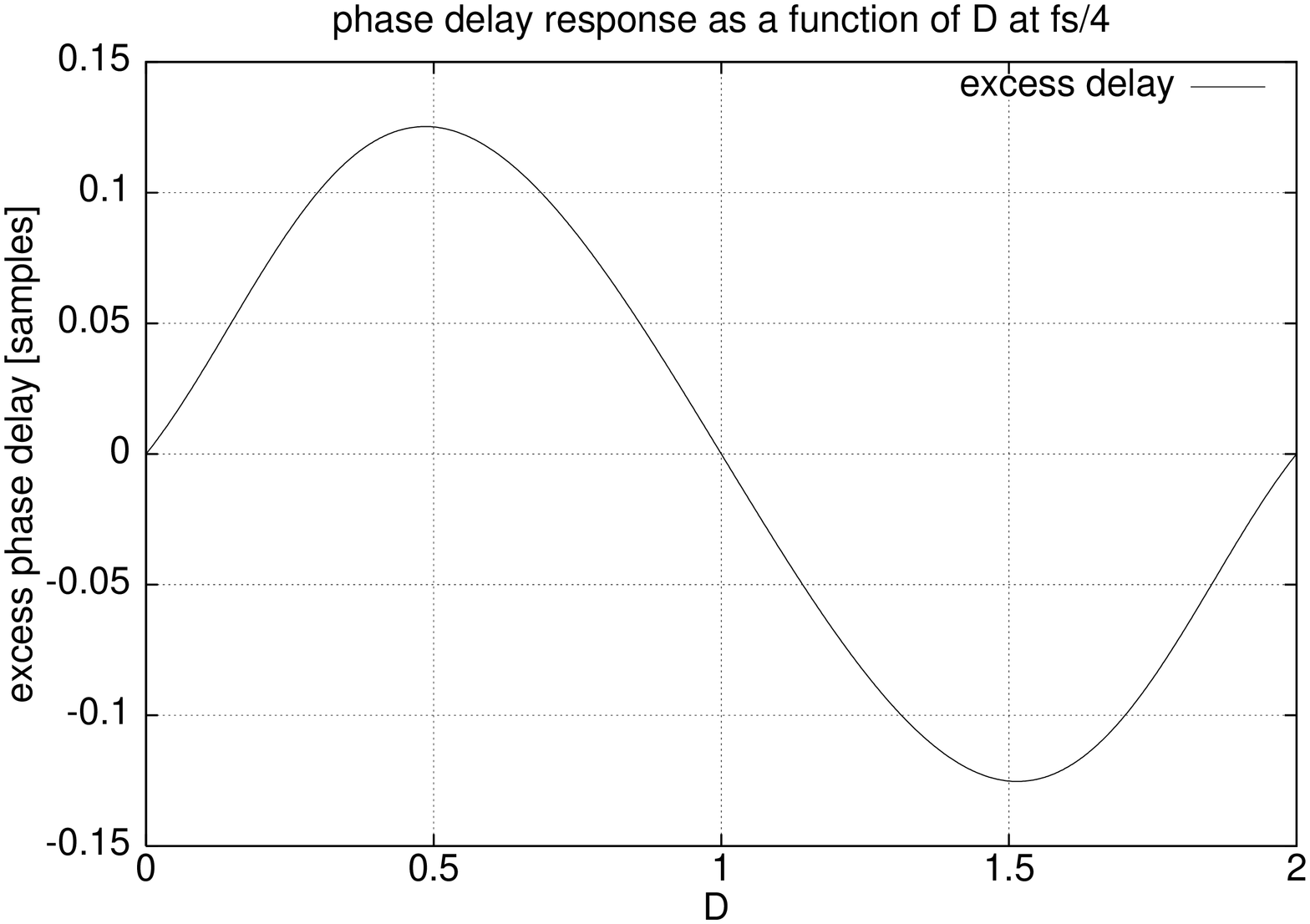,width=11.0cm,angle=-90}}
\centerline{(b)}
\caption{Magnitude (a) and excess phase-delay (b) responses for a quadratic Lagrange interpolator as functions of the parameter $D = 1 - d$ at frequency $f=F_s/4$}
\label{quaddel}
\end{figure}

\clearpage

\begin{figure}[ht]
\centerline{\psfig{file=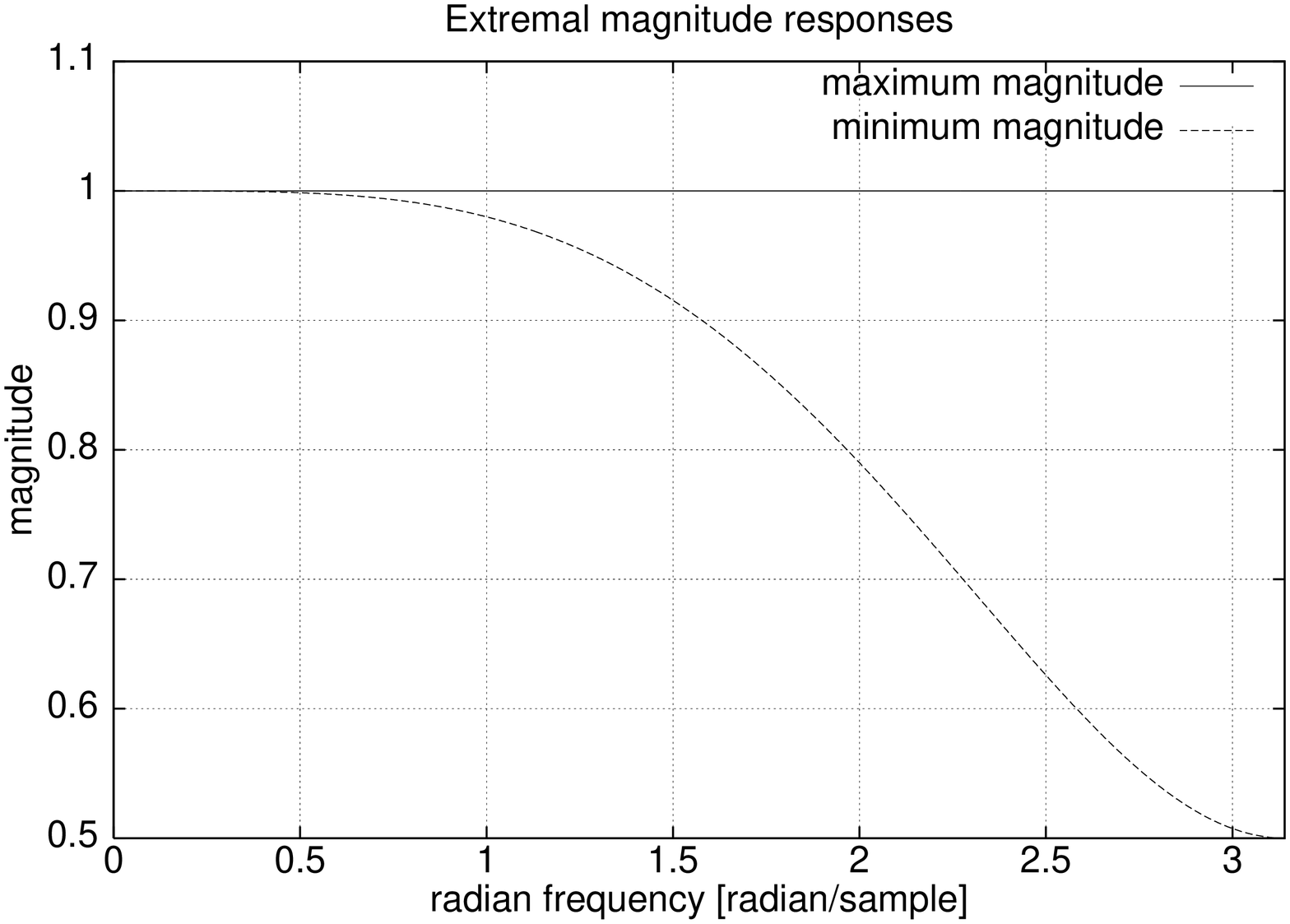,width=11.0cm,angle=-90}}
\centerline{(a)}
\vspace{0.1cm}
\centerline{\psfig{file=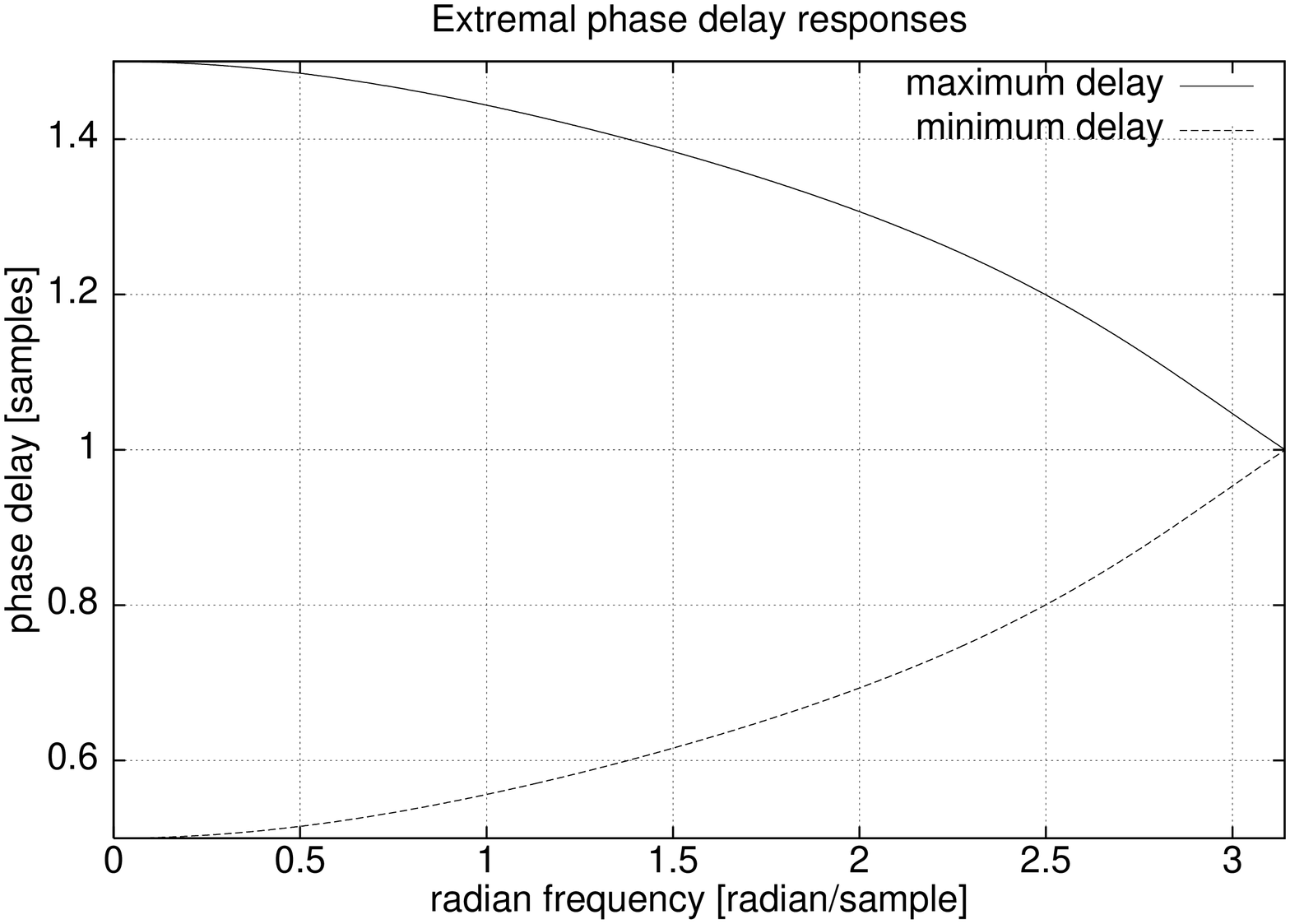,width=11.0cm,angle=-90}}
\centerline{(b)}
\caption{Extremal magnitude (a) and phase delay (b) responses for a quadratic Lagrange interpolator}
\label{quadmagdel}
\end{figure}

\clearpage

\begin{figure}[hbt]
\centerline{\psfig{file=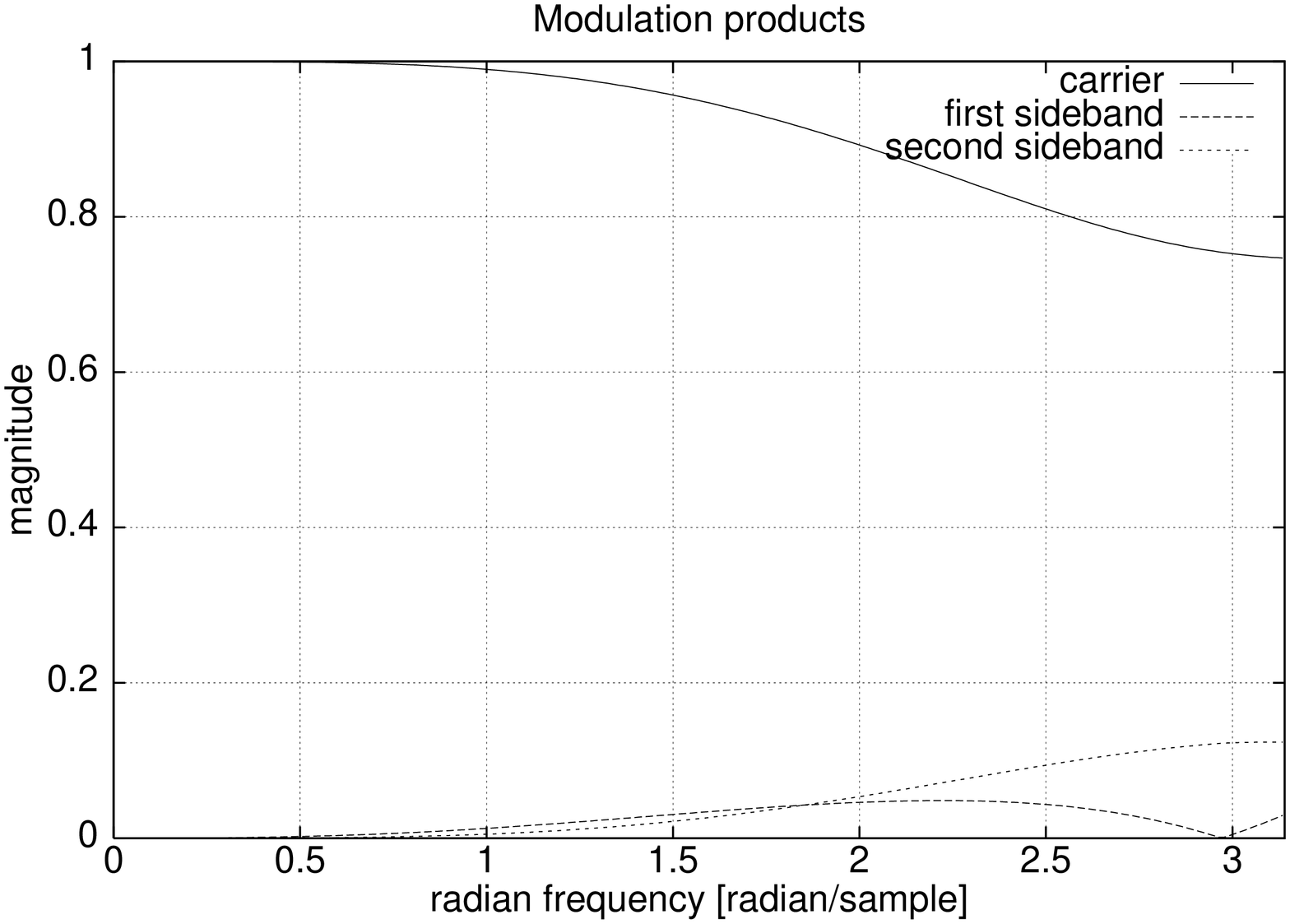,width=11.0cm,angle=-90}}
\caption{Carrier and side modulation products as a function of carrier frequency }
\label{components}
\end{figure}

\clearpage

\begin{figure}[hbt]
\centerline{\psfig{file=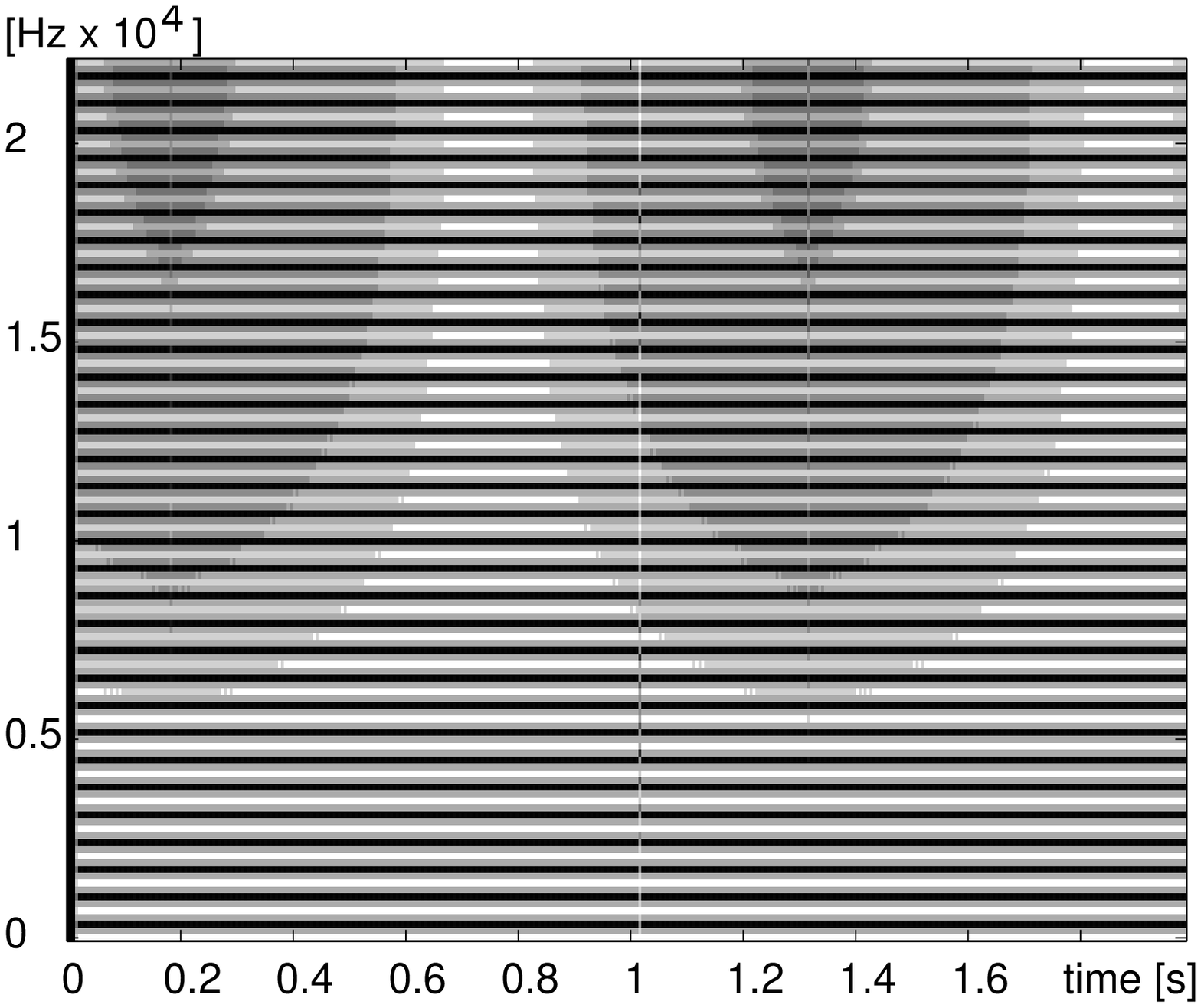,width=11.0cm}}
\centerline{(a)}
\vspace{0.1cm}
\centerline{\psfig{file=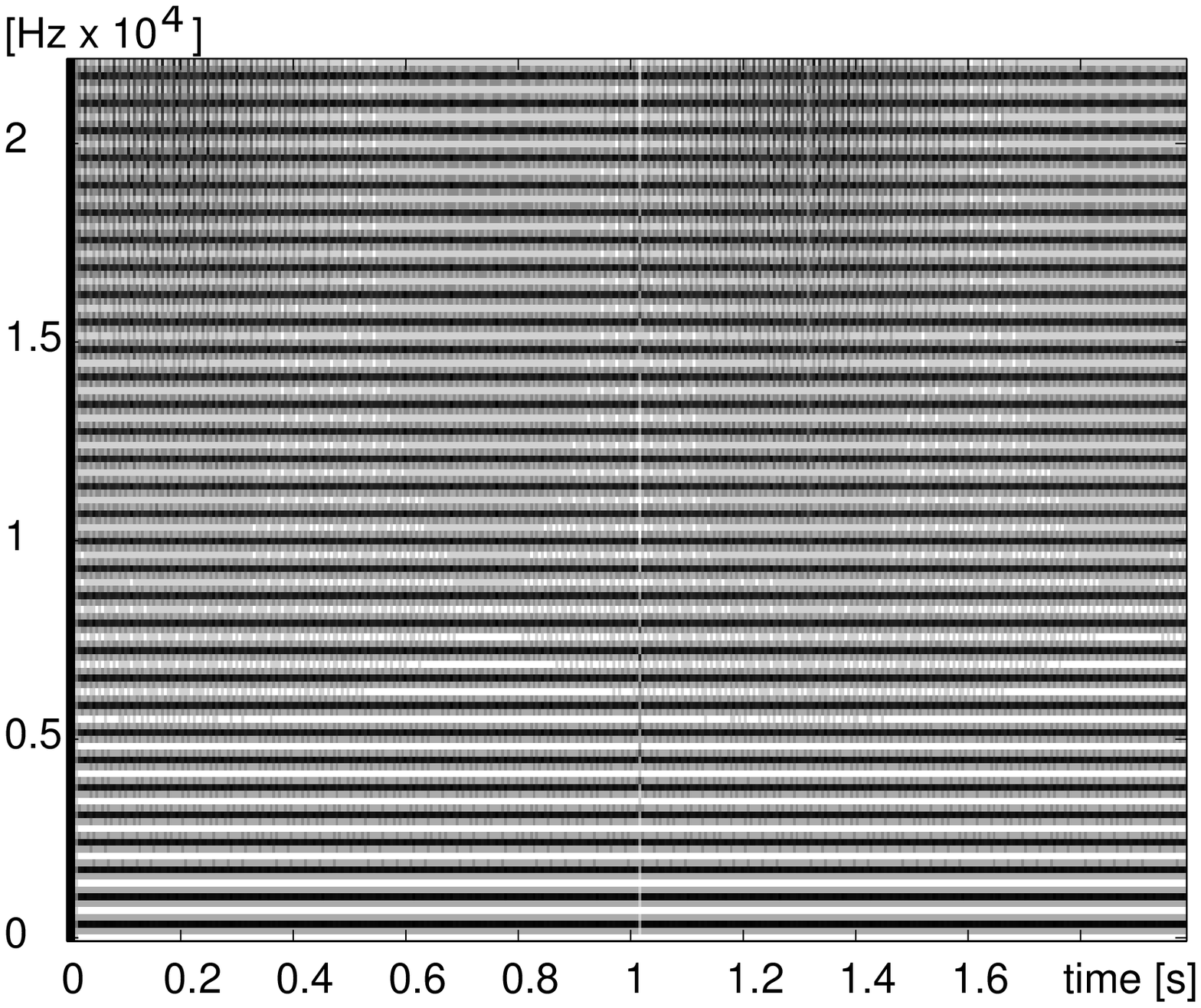,width=11.0cm}}
\centerline{(b)}
\caption{Sonogram of the  response of a non-modulated (a) and modulated (b) FIR delay line with linearly-increasing length to a pulse train. Hanning windows of 256 samples are used in analysis. Magnitude (in dB) is smaller where the points are darker.}
\label{noisyramp}
\end{figure}

\clearpage

\begin{figure}
{\tt
\begin{verbatim}

delay = Q/P * lenbuf; %delay in samples
framelen = floor(delay); 
for n=1:nframes
  bufout = resample(buffer, P, Q)'; %read
  output = [[output, bufout]];
  fwrite(fid_out, bufout, 'int16');
  bufin = fread(fid_in, framelen, 'int16');
  buffer = resample(bufin, Q, P); %write
end
\end{verbatim}
}
\caption{MATLAB code for a frame-based realization of the FAD line}
\label{framecode}
\end{figure}

\clearpage

\lungo{
\begin{figure}[htb]
\centerline{\psfig{file=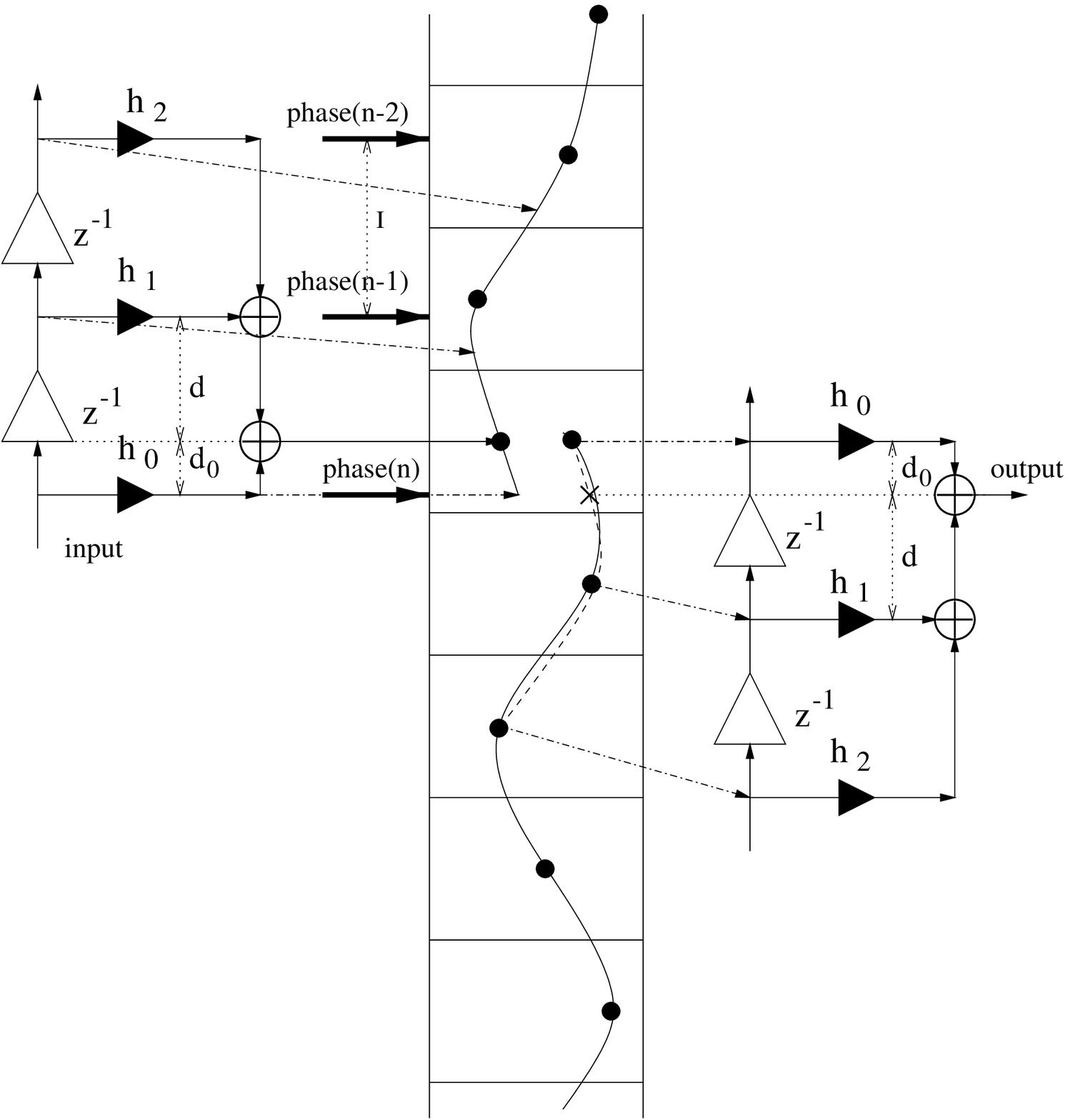,width=11.0cm}}
\caption{Interpolated read and write access to a circular buffer}
\label{buffer}
\end{figure}
}

\clearpage

\begin{figure}
\begin{tt}
\begin{verbatim}
loop
     fph = floor(phase);
     output = interpolated_read(table[fph],
             table[fph+1], ...);
     ph = (phase_old + 1) MOD length_table;
     while (ph <= fph) {
       table[ph] = interpolated_write(
         ..., table[phase_old], input);
       ph = (ph + 1) MOD length_table;
     }
     phase_old = fph;
     phase = (phase + Increment);
     if (phase > length_table)
          phase = phase - length_table;
endloop
\end{verbatim}
\end{tt}
\caption{Pseudo-code for a sample-by-sample realization of the FAD line}
\label{samplecode}
\end{figure}

\clearpage

\lungo{
\begin{figure}[htb]
\centerline{\psfig{file=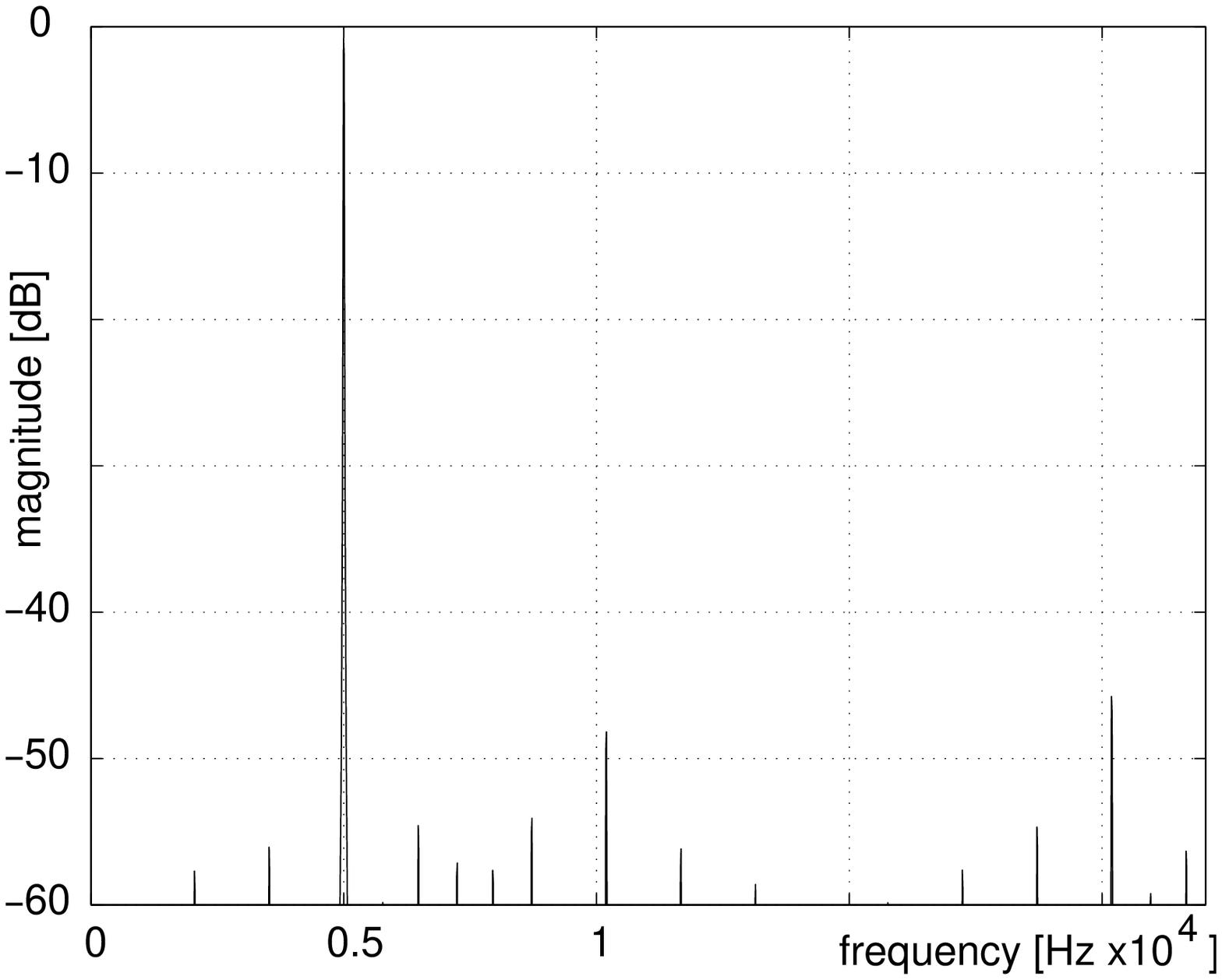,width=11.0cm}}
\caption{Magnitude spectrum of the output signal of a FAD line, where the
input signal is a sine wave at $5000 \hbox{Hz}$, the delay is $0.74378
\hbox{s}$, the sampling rate is $44.1 \hbox{kHz}$ and the buffer is $44100-\hbox{samples}$ long.  }
\label{outi}
\end{figure}
}

\clearpage

\begin{figure}[htb]
\centerline{\psfig{file=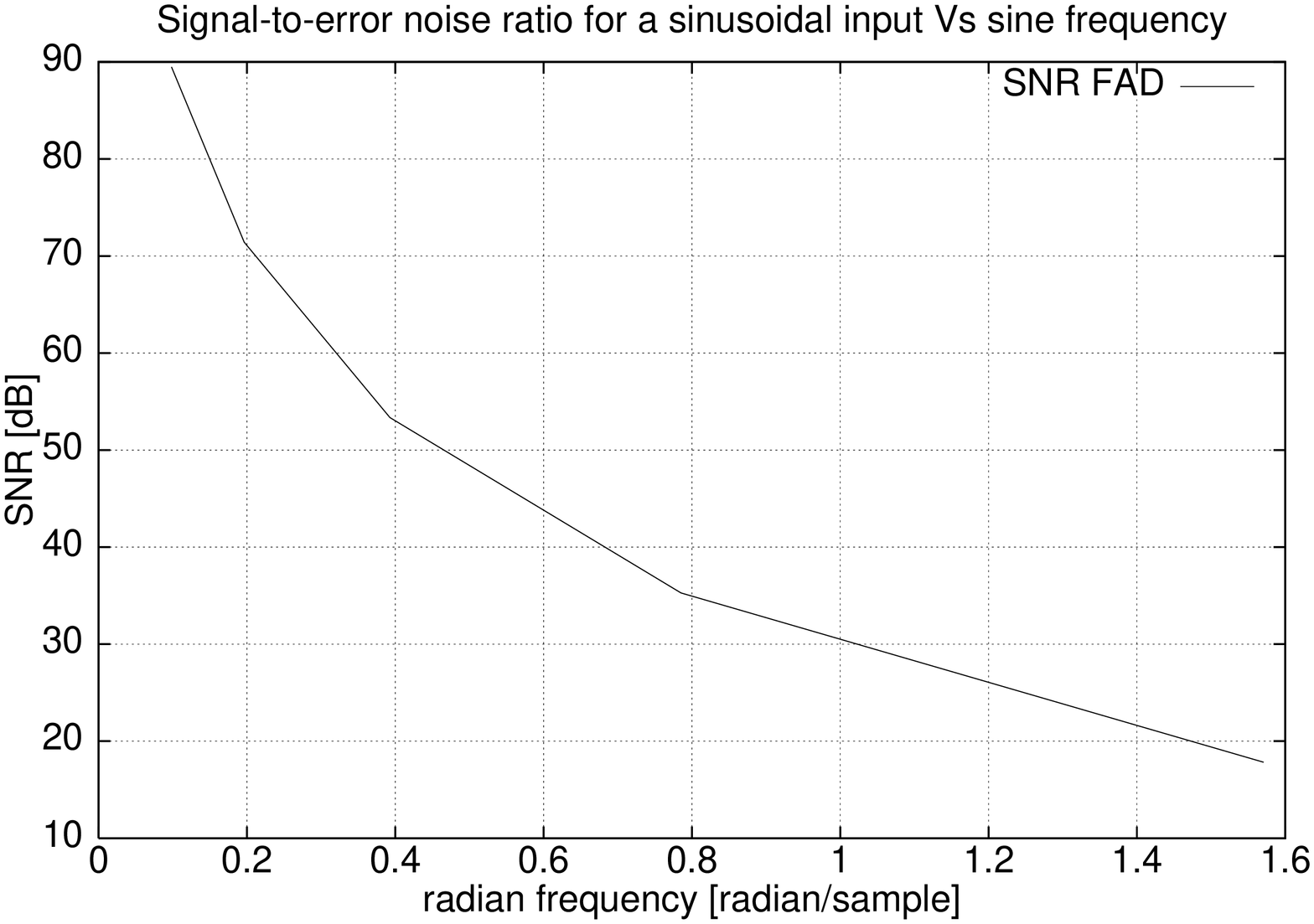,width=11.0cm,angle=-90}}
\caption{Experimental signal-to-error noise ratio Vs. sine frequency for the
FAD line  with quadratic interpolation. 
The input frequencies are such that the interpolation phase is a multiple of $2\pi$.}
\label{SNRcompared}
\end{figure}


\begin{figure}[hbt]
\centerline{\psfig{file=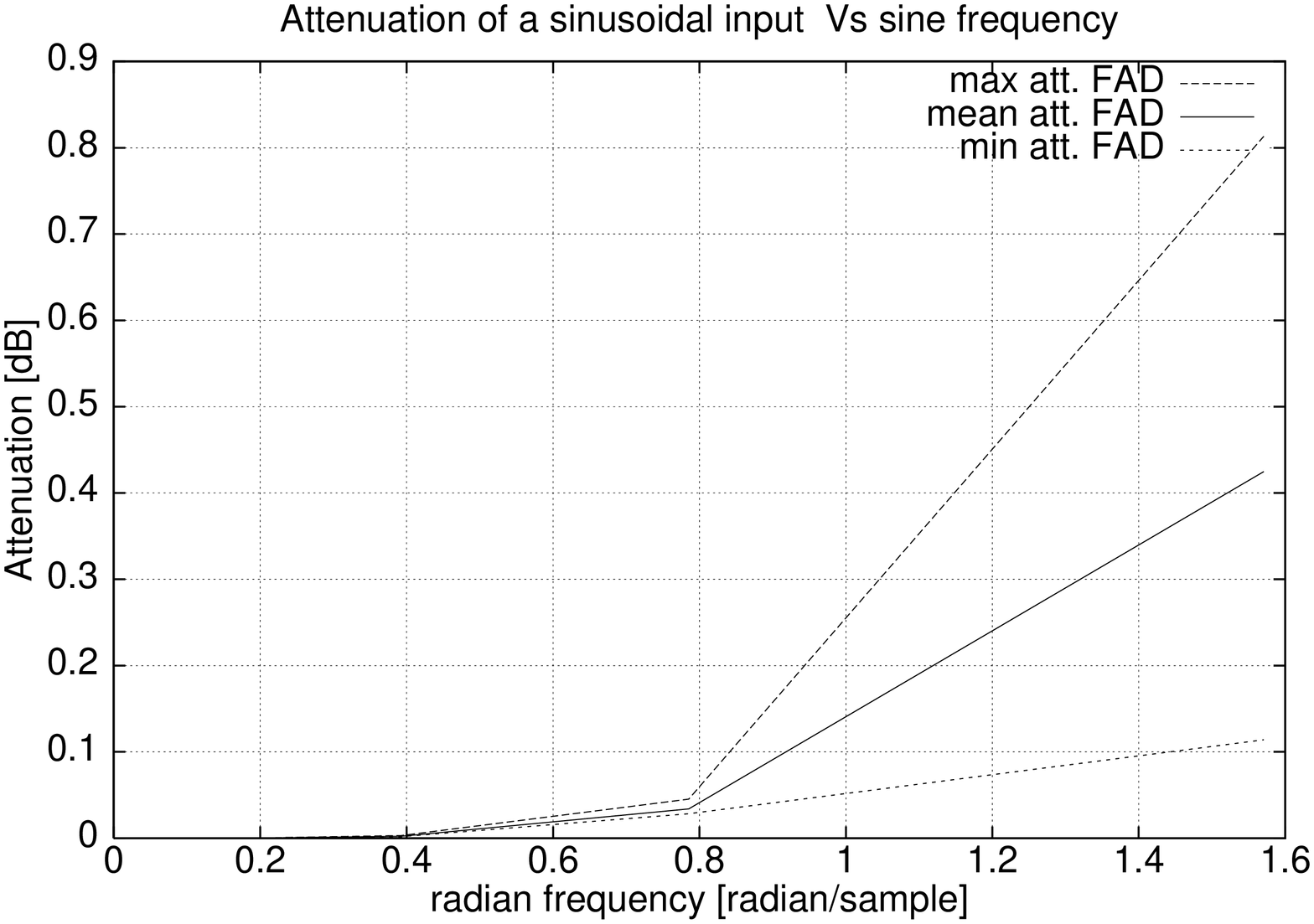,width=11.0cm,angle=-90}}
\caption{Attenuation of a sinusoidal input Vs. sine frequency for the
quadratically-interpolated FAD line. Sampling rate is set equal to the buffer length and delay is set to $D = 2/3 \hbox{s}$}
\label{attcompared}
\end{figure}

\begin{figure}[hbt]
\centerline{\psfig{file=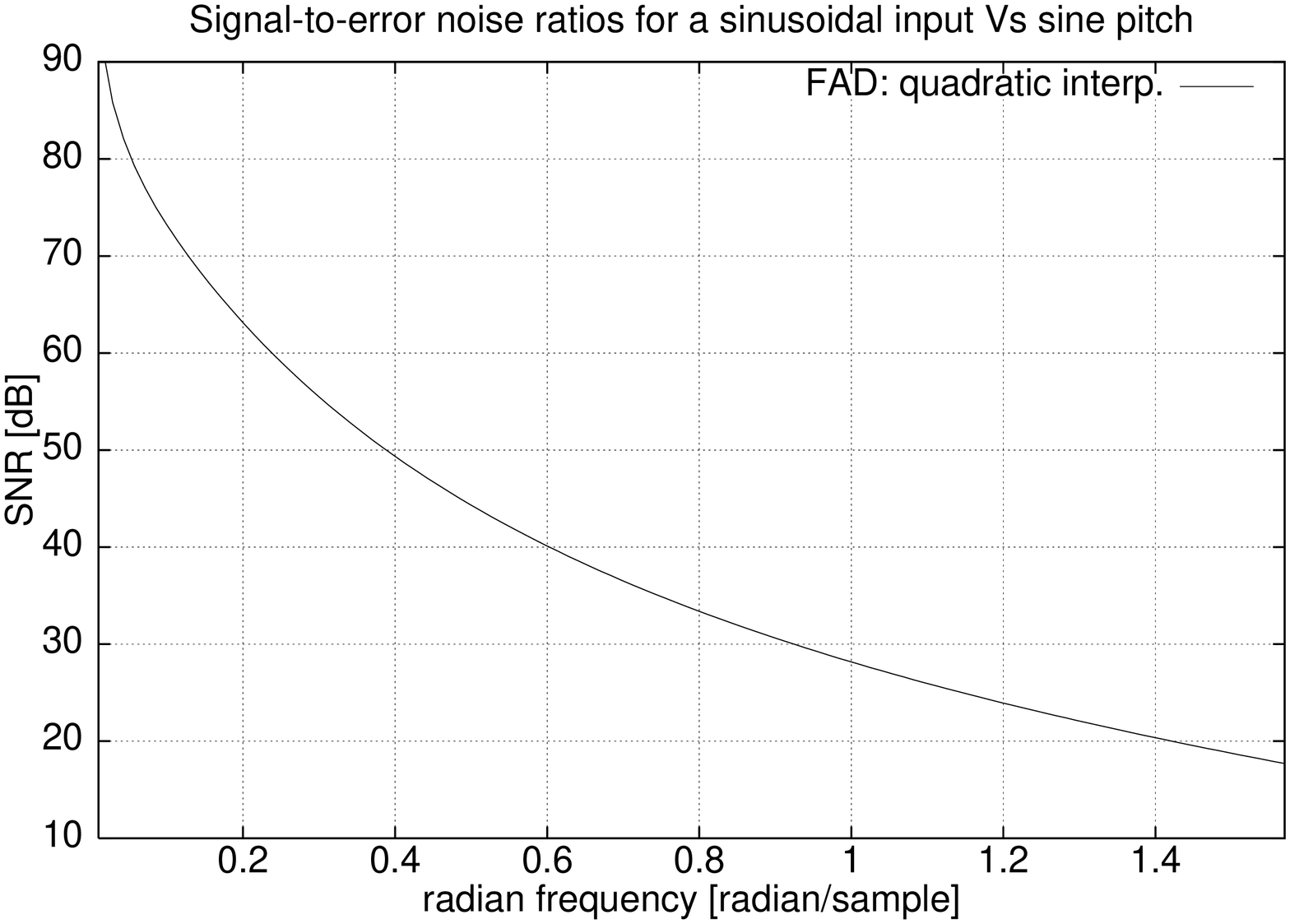,width=11.0cm,angle=-90}}
\caption{Signal-to-error noise ratio for the quadratic FAD line }
\label{snrPM}
\end{figure}

\clearpage

\begin{figure}[ht]
\centerline{\psfig{file=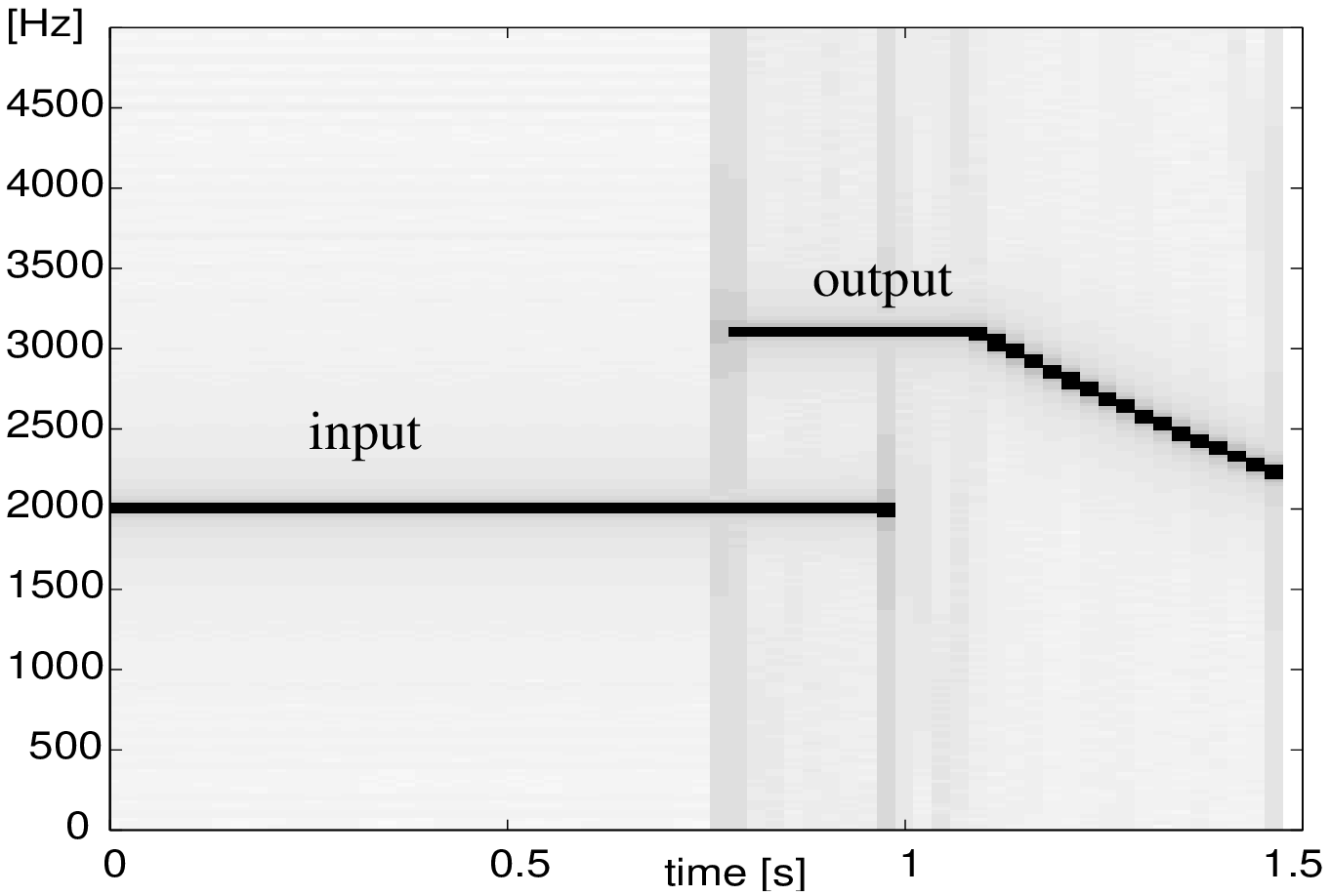,width=10.0cm}}
\caption{FAD line: delay ramp from 0.99 s to 0.5 s in 1.11 s; 1 s of sinusoidal input}
\label{gliss}
\end{figure}

\begin{figure}[ht]
\centerline{\psfig{file=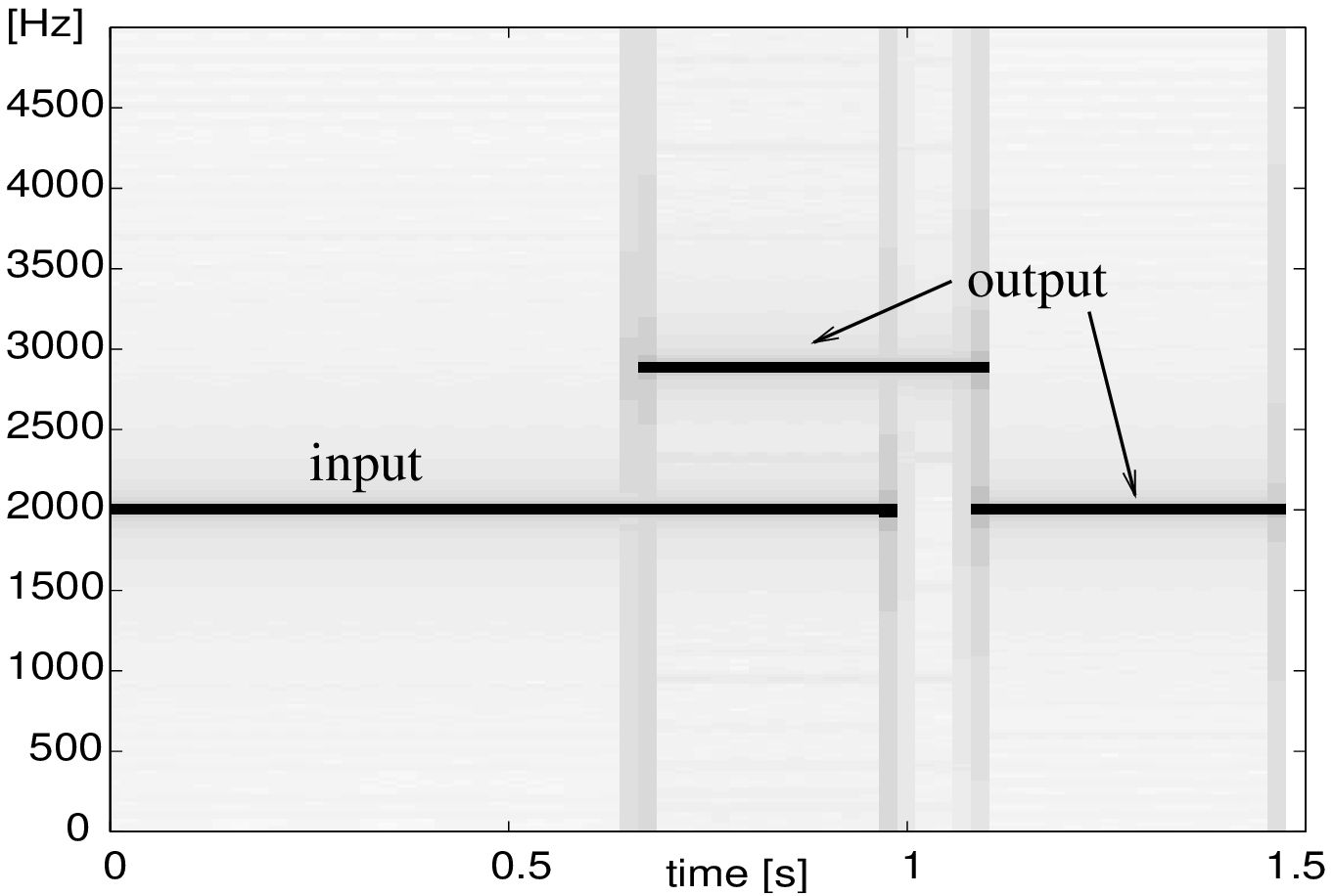,width=10.0cm}}
\caption{FIR line: delay ramp from 0.99 s to 0.5 s in 1.11 s; 1 s of
sinusoidal input} 
\label{glissf}
\end{figure}

\clearpage

\begin{figure}[ht]
\centerline{\psfig{file=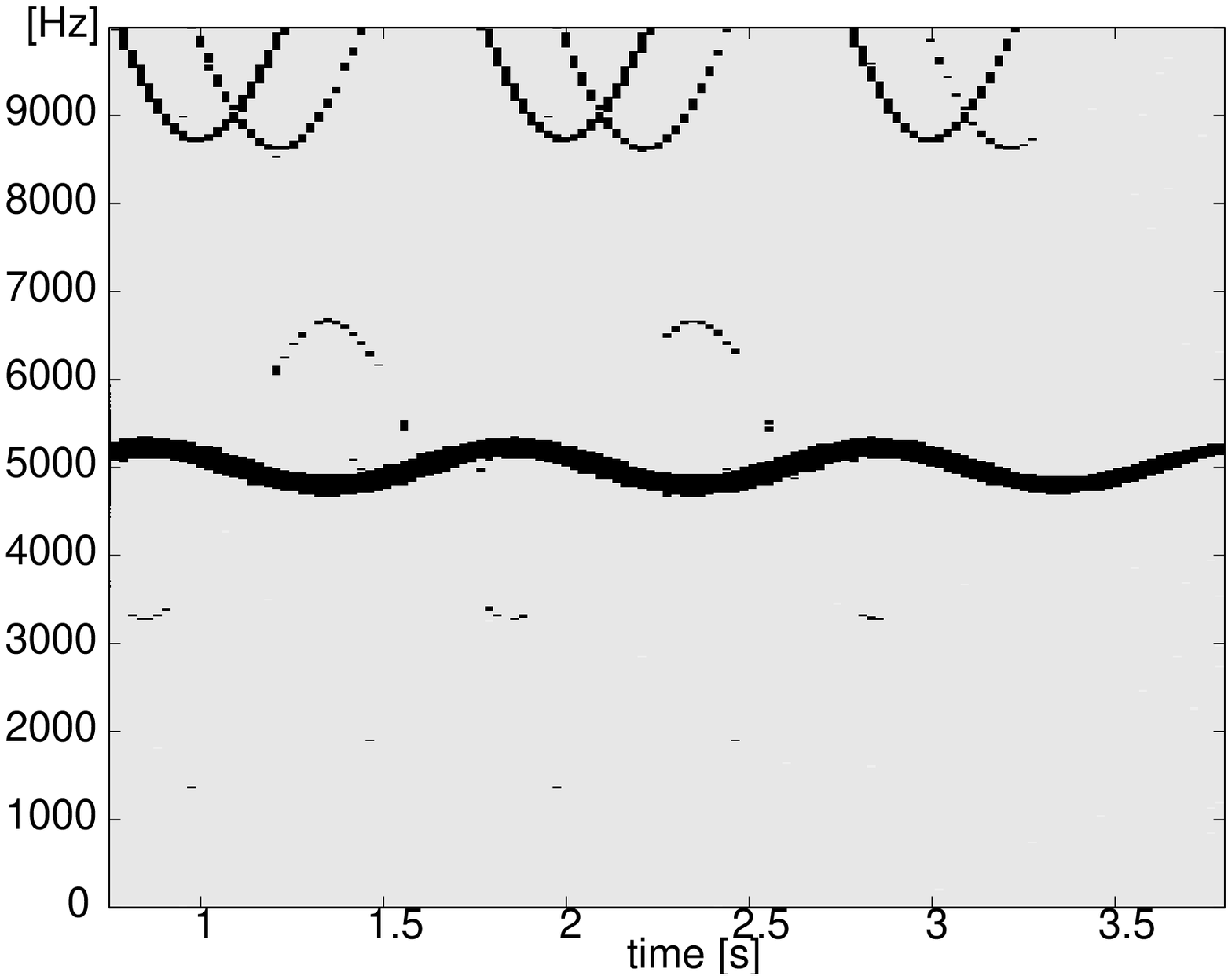,width=10.0cm}}
\caption{FAD line with quadratic interpolation: delay-length vibrato}
\label{vibrfa}
\end{figure}

\clearpage 

\begin{figure}[ht]
\centerline{\psfig{file=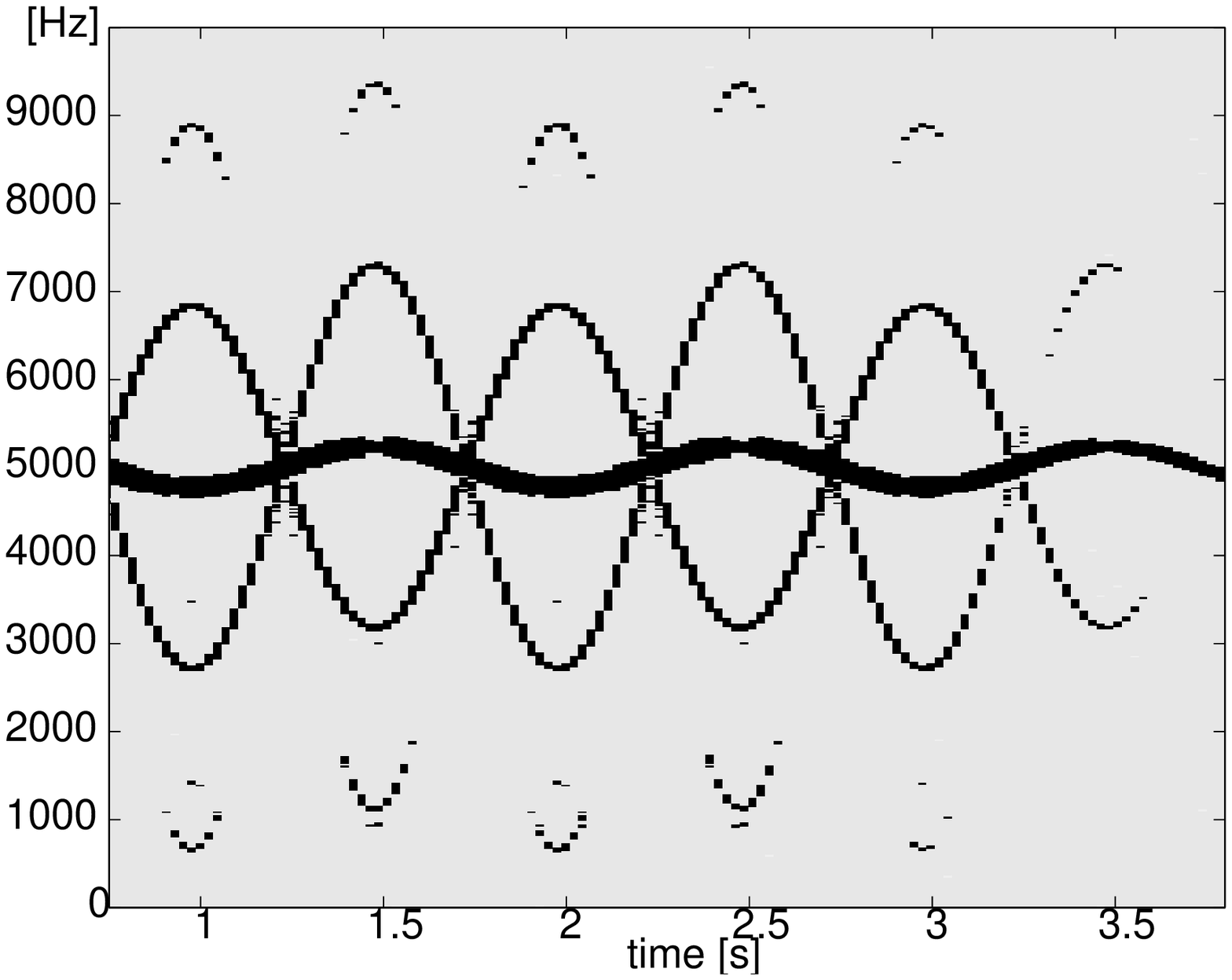,width=10.0cm}}
\caption{FIR line with quadratic interpolation: delay-length vibrato} 
\label{vibrfir}
\end{figure}

\clearpage
\begin{figure}
\vspace{3cm}
\centerline{\psfig{file=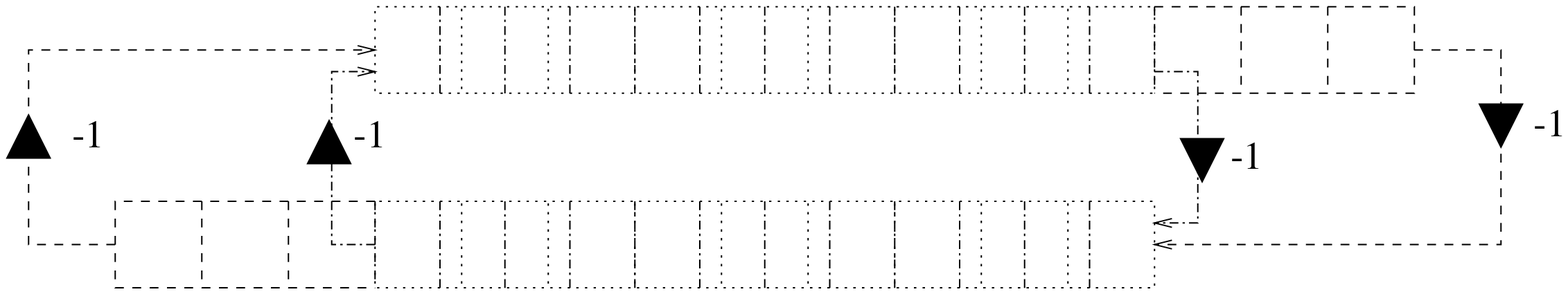,width=16.0cm,height=4cm}}
\caption{Waveguide model of an ideal string. A sudden pitch lowering is depicted for the FIR implementation of delay lines (dashed line) and for the FAD implementation of delay lines (dash-dotted line). The string before the pitch transition is represented by the dotted line.}
\label{waveguidestring}
\end{figure}

\clearpage

\begin{figure}[hb]
\centerline{\psfig{file=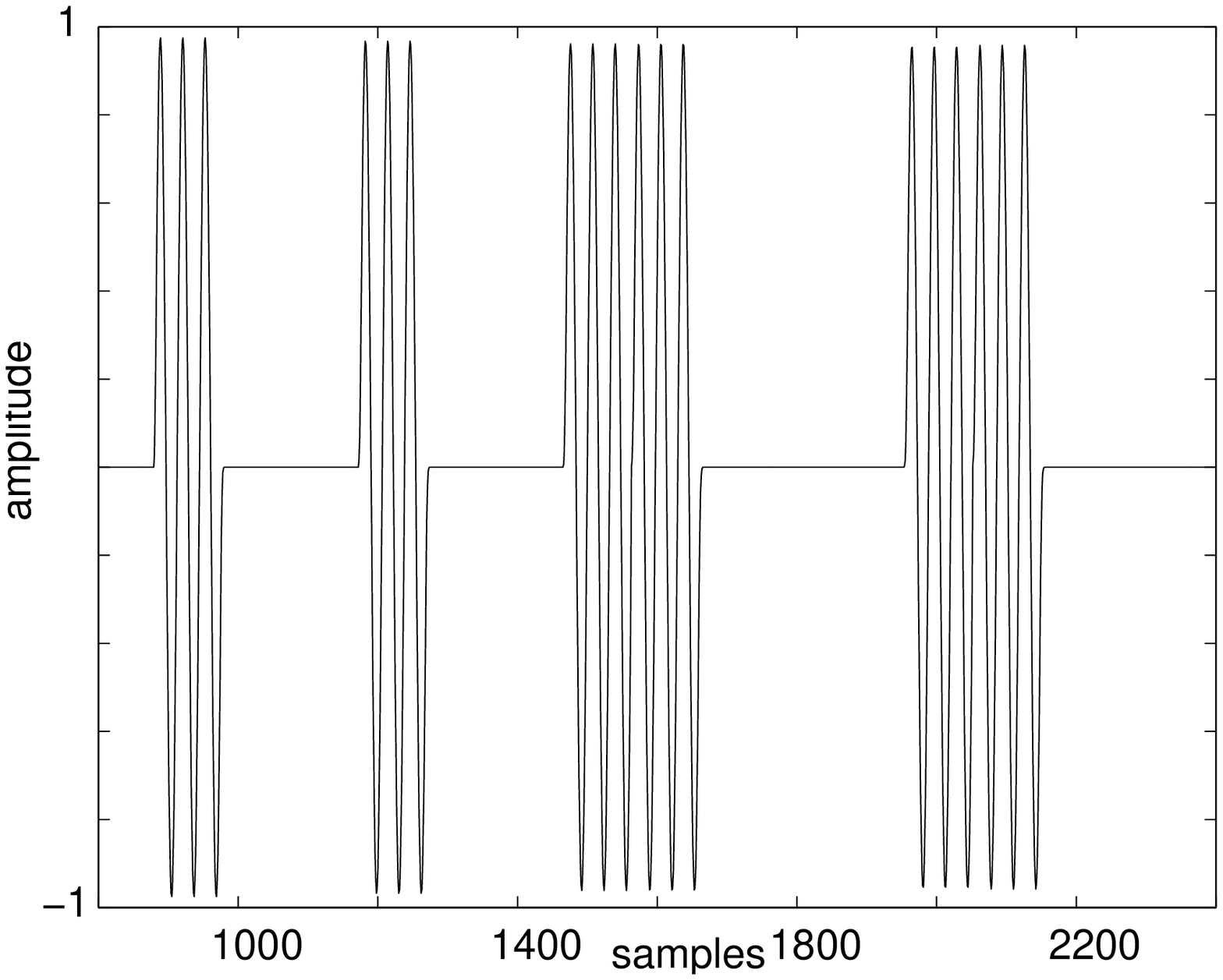,width=10.0cm,height=7.0cm}}
\caption{Note transition waveform: FIR line without erase after read}
\label{ntfirne}
\end{figure}

\begin{figure}[ht]
\centerline{\psfig{file=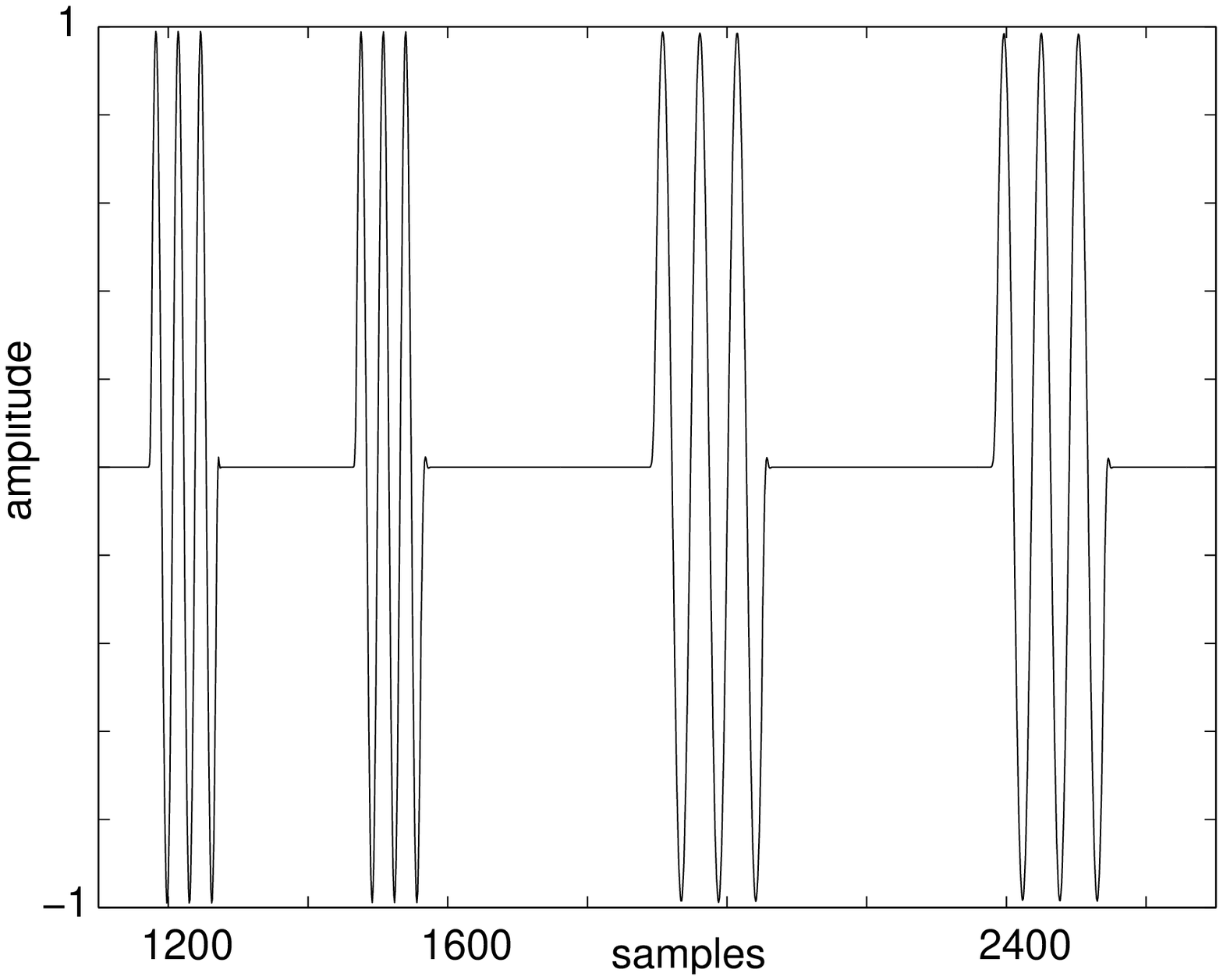,width=10.0cm,height=7.0cm}}
\caption{Note transition waveform: FAD line}
\label{ntfad}
\end{figure}

\Comment{\begin{figure}[hb]
\psfig{file=pitch3.wav.ps,width=10.0cm,height=5.0cm,angle=-90}
\caption{Note transition waveform: FIR line with erase after read}
\label{ntfire}
\end{figure}
}
\clearpage
\Comment{
\begin{figure}[htb]
\centerline{\psfig{file=pitch1.son.ps,height=4cm}}
\caption{Note transition sonogram: FAD line}
\label{ntfadson}
\end{figure}

\begin{figure}[htb]
\centerline{\psfig{file=pitch2.son.ps,height=4cm}}
\caption{Note transition sonogram: FIR line without erase after read}
\label{ntfirneson}
\end{figure}

\begin{figure}[htb]
\centerline{\psfig{file=pitch3.son.ps,height=4cm}}
\caption{Note transition sonogram: FIR line with erase after read}
\label{ntfireson}
\end{figure}

\clearpage
}
\begin{figure}[htb]
\centerline{\psfig{file=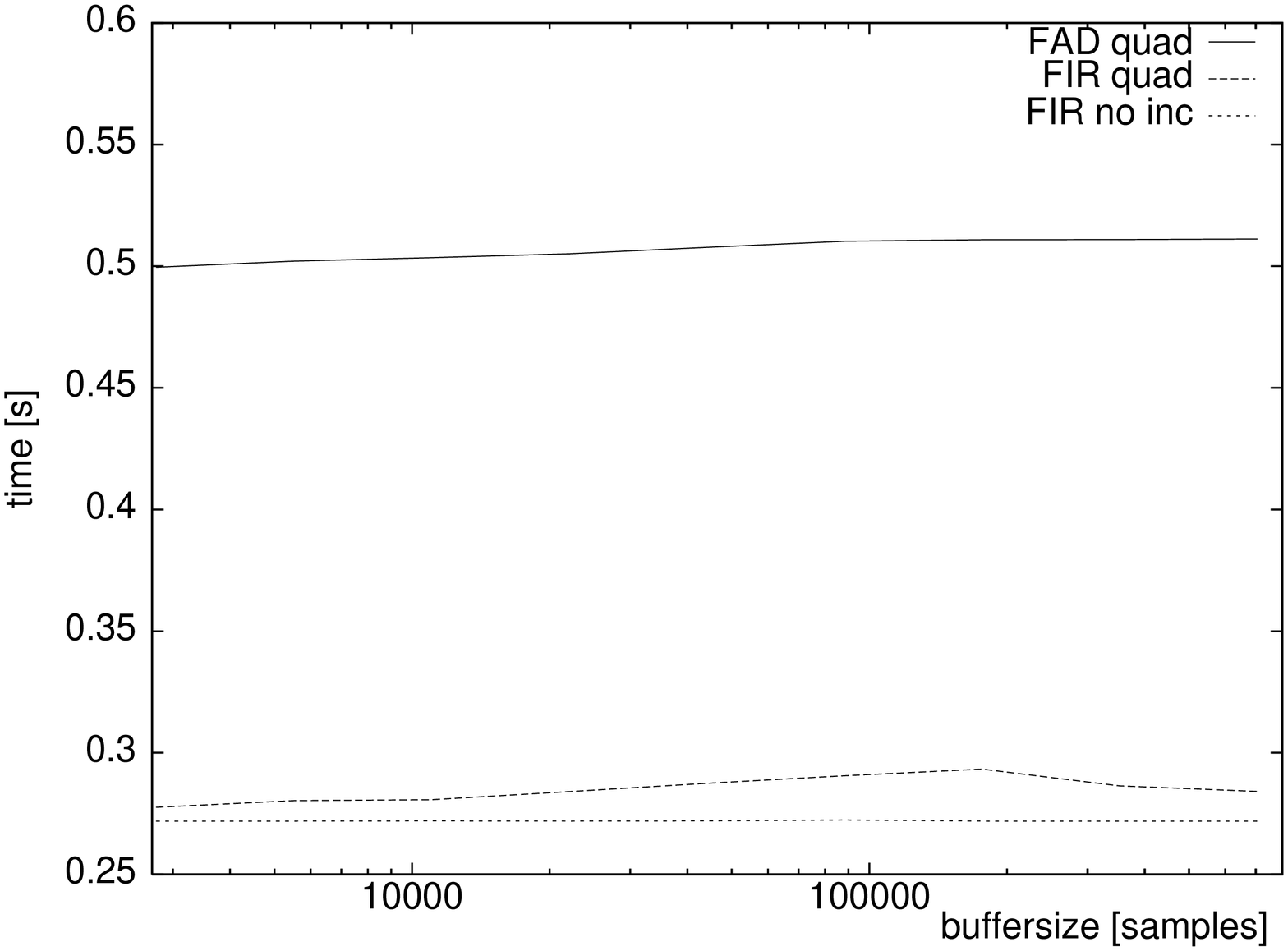,width=11.0cm,angle=-90}}
\caption{Performance on an AMD-K6 of the quadratically-interpolated FAD line, FIR line, and  FIR line with no phase increment, as a function of buffer size.}
\label{benchk6}
\end{figure}

\clearpage

\listoftables

\clearpage

\listoffigures

\fi

\clearpage

\begin{biography}{Davide Rocchesso} received the {\it Laurea in Ingegneria Elettronica} degree from the University of Padova in 1992, and the Ph.D. degree from the same university in 1996. His Ph.D. research involved the design of structures and algorithms based on feedback delay networks for sound processing applications. 
In 1994 and 1995 he was a visiting scholar at the Center for Computer Research in Music and Acoustics (CCRMA) at Stanford University. Since 1991 he has been collaborating with the {\it Centro di Sonologia Computazionale} (CSC) at the University of Padova as a researcher and a live-electronic designer. Since march 1998 he has been with the {\it Dipartimento Scientifico e Tecnologico} at the University of Verona, as an Assistant Professor. His main interests are in sound processing, physical modeling, sound reverberation and spatialization, multimedia systems. His home page on the web is http://www.sci.univr.it/\~{}rocchess.
\end{biography}

\end{document}